\documentclass[aps,prd,twocolumn,superscriptaddress]{revtex4-1}

\usepackage{longtable}
\usepackage{grffile} 
\usepackage{graphics}
\usepackage{graphicx} 
\usepackage{amsmath}    
\usepackage{hyperref}   
\usepackage{subfigure}  
\usepackage{epsfig}
\usepackage{epstopdf}
\usepackage{color}

\newcommand{\pup}{p^{\uparrow}}

\newcommand{\kp}{k_{\perp}}
\newcommand{\kpv}{\vec{k}_{\perp}}
\newcommand{\kpvhat}{\hat{\vec{k}}_{\perp}}
\newcommand{\xprec}{x_{parton}^{rec}}
\newcommand{\gauskp}{\frac{e^{-\kp^2/<\kp^2>}}{\pi <\kp^2>}}

\newcommand{ \gammapup }{ \gamma^{*}+\pup\rightarrow h_{1}+h_{2}+X }

\newcommand{ \phase }{ dz_{h1}dz_{h2}d^{2}p_{h1\perp}d^{2}p_{h2\perp}  }

\newcommand{\dsigminus}{d\sigma^{\uparrow}(\phi_{kS},k_{T})-d\sigma^{\downarrow}(\phi_{kS},k_{T})}
\newcommand{\dsigsum}{d\sigma^{\uparrow}(\phi_{kS},k_{T})+d\sigma^{\downarrow}(\phi_{kS},k_{T})}

\begin{document}

\title{Accessing the Gluon Sivers Function at a future Electron-Ion Collider}

\author{L. Zheng}\email{zhengliang@cug.edu.cn}
\affiliation{School of Mathematics and Physics, China University of Geosciences (Wuhan),
	\break Wuhan 430074, China}
\affiliation{Key Laboratory of Quark and Lepton Physics (MOE) and \break Institute
	of Particle Physics, Central China Normal University,\break Wuhan 430079, China}

\author{E.C. Aschenauer}\email{elke@bnl.gov}
\affiliation{Physics Department, Brookhaven National Laboratory, \break Upton, NY 11973, U.S.A.}
\author{J.H. Lee}\email{jhlee@bnl.gov}
\affiliation{Physics Department, Brookhaven National Laboratory, \break Upton, NY 11973, U.S.A.}
\author{Bo-Wen Xiao}\email{bxiao@mail.ccnu.edu.cn}
\affiliation{Key Laboratory of Quark and Lepton Physics (MOE) and \break Institute
of Particle Physics, Central China Normal University,\break Wuhan 430079, China}
\author{Zhong-Bao Yin}\email{zbyin@mail.ccnu.edu.cn}
\affiliation{Key Laboratory of Quark and Lepton Physics (MOE) and \break Institute
of Particle Physics, Central China Normal University,\break Wuhan 430079, China}

\date{\today}

\begin{abstract}
In this work, we present a systematic study on the feasibility of probing the
largely unexplored transverse momentum dependent gluon Sivers function (GSF) in
open charm production, high p$_T$ charged di-hadron and di-jet production at a future high
energy, high luminosity Electron-Ion Collider (EIC). The Sivers function is a measure
for the anisotropy of the parton distributions in momentum space inside a
transversely polarized nucleon. It is proposed that it can be studied through
single spin asymmetries in the photon-gluon fusion subprocess in electron proton
collisions at the EIC. Using a well tuned Monte Carlo model for deep inelastic
scattering, we estimate the possible constraints of the GSF
from the future EIC data. A comparison of all the accessible measurements
illustrates that the di-jet channel is the most promising way to constrain the
magnitude of the GSF over a wide kinematic range.
\end{abstract}

\maketitle

\section{Introduction}
\label{sec:intro}
In recent years, an important forefront in hadron physics is to explore the 
2+1 dimensional partonic structure of nucleons by including information on the internal parton
transverse momentum and coordinate space distributions. These extensions have
significantly broadened our understanding to the nucleon structure  compared to
the 1d picture in the longitudinal momentum space. The transverse momentum structure
of nucleons is encoded in the transverse momentum dependent
(TMD)\cite{Perdekamp:2015vwa} parton distribution functions (TMDs),  which
contain information on both the longitudinal momentum fraction $x$ and the transverse 
(sometimes called intrinsic) motion $k_{\perp}$ of quarks and gluons inside a fast moving nucleon.

When including spin degrees of freedom, TMDs link information on the intrinsic
spin of a parton ($s_{q,g}$) and their transverse motion ($k_{\perp}^{q,g}$) to
the spin direction of the parent nucleon. At leading twist the most
general spin dependent TMD can be denoted by $f_1^{q,g}(x,k_{\perp}^{q,g};s_{q,g},S)$. 
At leading order, there are eight such combinations, yielding eight independent TMDs~\cite{Boer:2011fh}. The Sivers function $f^{\perp}_{1T}$~\cite{Sivers:1989cc}, 
which encapsulates the correlations between a parton's transverse momentum inside the 
proton and the spin of the proton, has received the widest attention both phenomenologically and
experimentally among all TMDs. It was found that the Sivers function is not
universal in hard-scattering processes~\cite{Efremov:2004tp}, which has its
physical origin in the rescattering of a parton in the color field of the
remnant of the polarized proton~\cite{Collins:2002kn}. Proving experimentally
the process dependence of the Sivers function is a very important test of the
non-Abelian nature of quantum chromo-dynamics (QCD) in TMD factorization.

Experimentally, the quark Sivers function has been measured in semi-inclusive
deep inelastic scattering (SIDIS) by the HERMES, COMPASS, and JLab Hall A
collaborations~\cite{Airapetian:2004tw, Alekseev:2008aa, Qian:2011py}. However,
due to the limited statistics precision and the narrow kinematic coverage of the
SIDIS data, only the valence quark Sivers function at moderate to high $x$ could
be constrained in phenomenological fits~\cite{Anselmino:2008sga}. The quark Sivers
function has also been studied in polarized proton-proton collisions by the STAR
and PHENIX collaborations at the Relativistic Heavy Ion Collider
(RHIC)~\cite{Adamczyk:2012xd,Adare:2010bd}. There are first indications both
from STAR through the $W$-boson measurement~\cite{Adamczyk:2015gyk} and COMPASS in
Drell-Yan (DY) production~\cite{Aghasyan:2017jop} for the non-university of the
Sivers function~\cite{Collins:2002kn} if measured in hadron-hadron collisions or
SIDIS, but the still challenging statistical precision limits a definite conclusion. 
Both STAR and COMPASS will soon increase the statistical precision of these measurements 
by including recent high statistics data. 
At the future high energy, high luminosity Electron-Ion Collider (EIC)~\cite{Accardi:2012qut}, 
the quark Sivers function can be well constrained over a very wide kinematic range ($x, Q^2, z$ and $p_T$) in SIDIS with exquisite precisions. It has been systematically investigated in the one and two
hadron final states in Ref.~\cite{Matevosyan:2015gwa} with a modified PYTHIA event generator
that includes the quark Sivers effect.

The gluon Sivers function (GSF), on the other hand, is barely known at the
current stage~\cite{Boer:2015vso}. Presently, the major theoretical
constraint for the GSF comes from the Burkardt sum rule~\cite{Burkardt:2004ur},
which requires the total transverse momentum of all partons in a transversely
polarized nucleon to vanish. The only direct constrain of the GSF comes from the
left-right asymmetry $A_N$ data in $\pup p \rightarrow \pi^{0} X$ within
the so-called TMD generalized parton model (GPM) framework~\cite{D'Alesio:2015uta}. 
This analysis found that the gluon Sivers function is not large~\cite{Aschenauer:2015ndk}. 
However, the gluon Sivers obtained in the GPM may differ from the gluon Sivers 
function in the TMD framework~\cite{Boer:2015vso}. At this moment the only experimental 
constraint to the gluon Sivers function in the TMD framework comes from the recent SIDIS
measurement of high-$p_{T}$ hadron pairs off transversely polarized deuterons and protons at COMPASS~\cite{Adolph:2017pgv}. This analysis found that the gluon Sivers asymmetry is 
negative at large $x_B$ within statistical uncertainties. Interestingly, 
this finding is in qualitative agreement with results from the calculation based 
on the light-cone spectator model\cite{Lu:2016vqu}.

Accounting for the different gauge link structures involved in deep-inelastic scattering (DIS) 
and hadronic collisions, the gluon Sivers function is expected to be process dependent. A test of 
this non-universality of the gluon Sivers function is of equal importance as for the quarks and is currently not validated.
Similar to the sign change of the quark Sivers function, the gluon Sivers function accessed 
in $e \pup \rightarrow e' c\bar{c} X$ is also predicted to be related to that in 
$\pup p\rightarrow \gamma \gamma X$ by an overall sign change 
$f_{1T}^{g}[e \pup \rightarrow e' c \bar{c} X]=-f_{1T}^{g}[\pup p\rightarrow \gamma \gamma X]$ 
as shown in~\cite{Boer:2016fqd}. Consequently, the study of the gluon Sivers function at
an EIC will provide a unique test of the fundamental non-perturbative QCD effects through 
complementary information to the proposed gluon Sivers function observables at RHIC and LHC~\cite{Brodsky:2012vg,Kikola:2017hnp}. 
In addition, as pointed out in Ref.~\cite{Dominguez:2010xd}, there are two different 
type of gluon TMDs, namely, the Weizs\"{a}cker-Williams and the dipole gluon distribution. 
This is a direct consequence of the different gauge link dependences. By comparing the gluon 
Sivers functions extracted from DIS and $pp$ collisions, one can test this gauge link dependence 
since the Weizs\"{a}cker-Williams and dipole type T-odd gluon TMDs are expected to behave
differently \cite{Boer:2015vso, Boer:2016fqd}. 

In DIS, the key to study the gluon Sivers function is to tag the Photon-Gluon
Fusion (PGF) subprocess. It has been shown in Ref.~\cite{Dominguez:2010xd,Boer:2010zf} that
the gluon transverse momentum distribution can be mapped through quark-antiquark
jet correlations for the PGF subprocess $\gamma^{*}g\rightarrow q\bar{q}$.
Ref.~\cite{Boer:2016fqd} suggests that the spin asymmetries measured in heavy
quark pair and di-jet production at an EIC can be used to study the Weizs\"{a}cker-Williams (WW)
gluon TMDs including the Sivers function. The open charm production in electron-proton
scattering $e \pup \rightarrow e' c \bar{c} X$ is argued to be an ideal probe to
tag the PGF process, and can be investigated at a future EIC. A model study has
been carried out in~\cite{Boer:2011fh} and the related experimental
considerations for tagging charm quark production through $D$-mesons in the
final state for the PGF subprocess are discussed in~\cite{Burton:2012ug}. 
In this paper, we will provide detailed information on EIC projections for open charm 
production with attainable experimental conditions. Alternative methods tagging the gluon
channel through the production of high-$p_T$ hadron pairs and di-jets are also
studied. The advantages and disadvantages of the different channels will be
discussed. Table~\ref{tab:varDef} shows the definitions of the kinematic
variables used in this paper.

\begin{longtable*}[H]{ll}
\caption{ Kinematic variables \label{tab:varDef} } \\ \hline \hline
$Q^{2}$	& Virtuality of exchanged photon  \\
$x_{B}$ & Bjorken $x$\\
$y$	& Energy fraction of virtual photon with respect to incoming electron \\
$W$	& Center of mass energy of the $\gamma^{*}p$ system\\
$x$	& Longitudinal momentum fraction of the quark/gluon from the polarized proton 
involved in the hard interaction\\
$z_{h,q}$	& Energy fraction of of a hadron or quark with respect to virtual 
photon in target rest frame \\
$k_\perp$ & Initial transverse momentum of gluons inside the proton in 
$\gamma^{*}p$ center of mass frame \\
$k_{1\perp}, k_{2\perp}$ &	Transverse momentum of the two outgoing partons in 
$\gamma^{*}p$ center of mass frame \\
$p_{h1\perp}, p_{h2\perp}$ &	Transverse momentum of the trigger/associate 
particle in $\gamma^{*}p$ center of mass frame \\
$p_{T}$ & Transverse momentum of final state hadron/jet with respect to virtual photon \\
$\eta$  & Pseudorapidity of final state hadron/jet in $\gamma^{*}p$ center of mass frame \\
$P_{T}$ & Transverse momentum scale of final state particle/jet pair with respect 
to virtual photon \\
$k_{T}$	& Vector sum of the transverse momentum for the final state hadron/jet pair 
in the final state \\
$\phi_{kS}$ & Sivers angle, the azimuthal angle difference of $k_{T}$ and the proton spin direction \\
$p_{T Lab}, p_{Lab}$ & Transverse momentum/momentum of final state hadron in the 
laboratory frame \\
$\eta_{Lab}$ & Pseudorapidity of the final state hadron/jet in the laboratory 
frame \\ \hline \hline
\end{longtable*}

The remainder of the paper is organized as follows: in Sec.~\ref{sec:gluon_sivers} we
discuss the theoretical framework used to build the connection of gluon Sivers function
and the size of the single spin asymmetry (SSA). The Monte Carlo setup is described in
Sec.~\ref{sec:mcsetup}. A detailed description of the results and their projected 
precision are presented in Sec.~\ref{sec:results}. We summarize in Sec.~\ref{sec:sum}.

\section{Single spin asymmetry (SSA) arising from the gluon Sivers effect}
\label{sec:gluon_sivers}

The Sivers function describes the distribution of unpolarized partons with
flavor $a$ inside a transversely polarized proton with mass $M_p$ and can be
expressed following the Trento convention in Ref.~\cite{Bacchetta:2004jz} as:

\begin{eqnarray}
\hat{f}_{a/\pup}(x, \kp) & = & f_{a/p}(x, \kp) \nonumber \\
 & + & \frac{1}{2} \Delta^{N}f_{a/\pup}(x,\kp) \vec{S} \cdot (\hat{\vec{P}}\times \kpvhat ).  
\label{eqn:sivers}
\end{eqnarray}

The first term represents the axially symmetric contribution from the
unpolarized parton distribution, while the second term
$\Delta^{N}f_{a/\pup}(x,\kp)$ generates a distortion away from the center in the
number density of unpolarized partons with an intrinsic transverse momentum
$\vec{k}_{\perp}$. The azimuthal dependence of this distortion is given by
$\vec{S} \cdot (\hat{\vec{P}}\times \kpvhat)$, where $\vec{P}$ and $\vec{S}$ are
the polarized proton three momentum and spin polarization vector, respectively.
The notation $\Delta^{N}f_{a/p^{\uparrow}}(x,\kp)$ is related to the Sivers
function denoted as $f^{\perp a}_{1T}(x, \kp)$ in the relation
$\Delta^{N}f_{a/p^{\uparrow}}(x,\kp) = -\frac{2\kp}{M_{p}}f^{\perp a}_{1T}(x,
\kp)$~\cite{Mulders:1995dh}.

The production of high transverse momentum charged hadron pairs or di-jets in DIS
through $\gamma^{*} g\rightarrow q \bar{q}$ is dominated by gluons, although it
may also have some contribution from the quark channel depending on the process
measured. The cross section can be calculated in an effective $k_{t}$
factorization framework at leading order as shown in
Ref.~\cite{Dominguez:2011wm}. If $k_1$ and $k_2$ are the four momenta of the
outgoing quarks, one can obtain the di-hadron cross section as a generalization
of the unpolarized case~\cite{Zheng:2014vka} with the transverse momentum
imbalance defined as $\kp=|\vec{k}_{1\perp}+\vec{k}_{2\perp}|$ and the
transverse momentum scale as $P_{\perp}=|\vec{k}_{1\perp}-\vec{k}_{2\perp}|/2$:

\begin{widetext}
\begin{eqnarray}
\frac{ d\sigma^{\gammapup}_{\textrm{tot}}}{\phase} = & \int^{1-z_{h2}}_{z_{h1}} 
\sum\limits_{q} dz_q \frac{z_q(1-z_q)}{z^{2}_{h2}z^{2}_{h1}}
d^{2}p_{1\perp}d^{2}p_{2\perp} \hat{f}_{g/p^{\uparrow}}(x,k_{\perp})  \nonumber \\ 
& \times \mathcal{H}_{\textrm{tot}}^{\gamma^{*} g\rightarrow q \bar{q}}(z_q,k_{1\perp},k_{2\perp}) D_{h1/q}(\frac{z_{h1}}{z_q},p_{1\perp})
D_{h2/\bar{q}}(\frac{z_{h2}}{1-z_q},p_{2\perp}), 
\label{eqn:pair_cs}
\end{eqnarray}
\end{widetext}

where $z_q$ is the momentum fraction of produced quark $q$ with respect to the incoming 
virtual photon and $\mathcal{H}_{\textrm{tot}}^{\gamma^{*} g\rightarrow q \bar{q}}$ gives the combined hard factor that incorporates both longitudinal part $\mathcal{H}^{\gamma^{*}_{L}g\rightarrow q \bar{q}}=\alpha_{s}\alpha_{em}e^{2}_{q}\frac{8\hat{s}Q^{2}}{(\hat{s}+Q^{2})^{4}}$ 
and transverse part $\mathcal{H}^{\gamma^{*}_{T}g\rightarrow q \bar{q}}=\alpha_{s}\alpha_{em}e^{2}_{q}\frac{\hat{s}^{2}+Q^{4}}{(\hat{s}+Q^{2})^{4}}(\frac{\hat{u}}{\hat{t}}+\frac{\hat{t}}{\hat{u}})$ of the virtual photon. Eq.~\ref{eqn:pair_cs} can be further simplified using the condition $\kp \ll P_{\perp}$
known as the correlation limit~\cite{Dominguez:2011wm}. 
Eq.~\ref{eqn:pair_cs} can thus be expressed as
\begin{widetext}
\begin{eqnarray}
\frac{ d\sigma^{\gammapup}_{\textrm{tot}}}{\phase} = & \int^{1-z_{h2}}_{z_{h1}} 
\sum\limits_{q} dz_q \frac{z_q^{2}(1-z_q)^{2}}{z^{2}_{h2}z^{2}_{h1}}
d^{2}p_{1\perp}d^{2}p_{2\perp}\alpha_s \alpha_{em} e_q^2\frac{[(z_q^2+(1-z_q)^2)(P_{\perp}^4+\epsilon_f^4)+8z_q(1-z_q)P_{\perp}^2\epsilon_f^2]}{(P_{\perp}^2+\epsilon_f^2)^4} \nonumber \\ 
& \times  \hat{f}_{g/p^{\uparrow}}(x,k_{\perp})  D_{h1/q}(\frac{z_{h1}}{z_q},p_{1\perp})
D_{h2/\bar{q}}(\frac{z_{h2}}{1-z_q},p_{2\perp}), 
\label{eqn:pair_cs_full}
\end{eqnarray}
\end{widetext}
in which $\epsilon_f^2$ is related to $Q^2$ as $\epsilon_f^2=z_q(1-z_q)Q^2$.
Choosing the center of mass frame of the exchanged virtual photon and the
proton, in which the proton beam with momentum $\vec{P}$ is moving in the $-z$
direction, one can obtain an explicit form of the mixed vector product in
Eq.~\ref{eqn:sivers} as $ \vec{S} \cdot (\hat{\vec{P}}\times \kpvhat) =
\sin(\phi_k-\phi_S)$ with $\phi_k$ being the azimuthal angle of $\kpv$. A
factorized Gaussian parameterization has been adopted for the transverse
momentum dependent unpolarized parton distribution function
$f_{g/p}(x,\kp)=f_{g/p}(x)\frac{e^{-\kp^2/<\kp^2>}}{\pi\langle\kp^2\rangle}$ and
fragmentation function $D(z,p_{\perp})=D(z)\frac{e^{-p_{\perp}^2/\langle
		p_{\perp}^2\rangle}}{\pi\langle p_{\perp}^2\rangle}$.

\begin{figure}
	\begin{center}
		\includegraphics[width=0.45\textwidth]{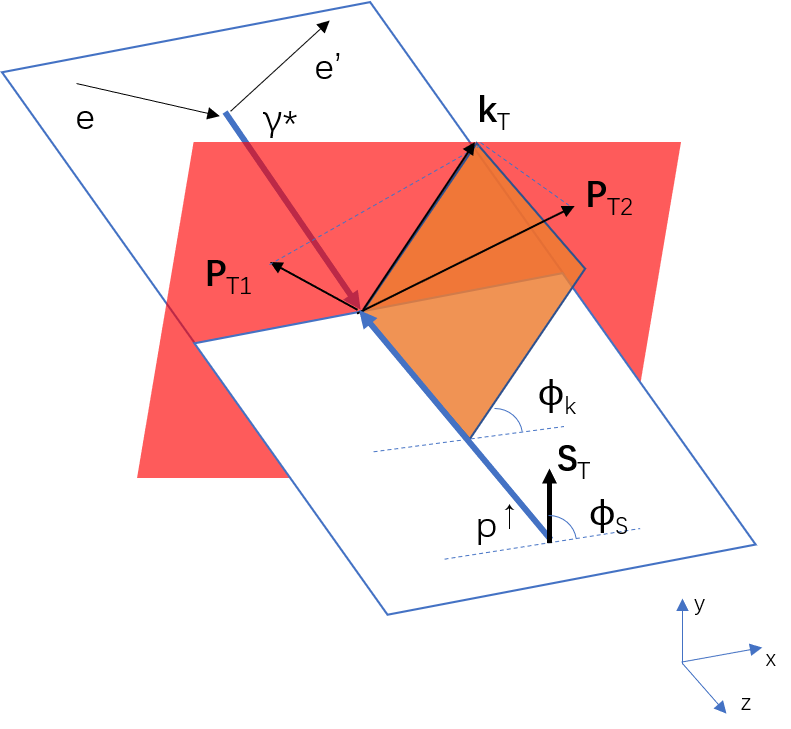}
	\end{center}
	\caption{(Color online) A schematic illustration of the kinematic variables involved 
		in this measurement.
	}
	\label{fig:layout}
\end{figure}

There exists a strong correlation between the kinematics of the back-to-back
hadron pair and its parent quarks. Therefore, one can use the following variables
measurable hadron level $P_{T}=|\vec{p}_{h1\perp}-\vec{p}_{h2\perp}|/2$ and
$k_{T}=|\vec{p}_{h1\perp}+\vec{p}_{h2\perp}|$ to access the underlying parton
kinematic variables $P_{\perp}$ and $\kp$. A schematic illustration of the
encoded kinematic variables is shown in Fig.~\ref{fig:layout}. In Sec.~\ref{sec:results},
a study, to which precision the measurable hadron level variables represent the
parton kinematics, will be presented. The GSF can be
studied in the single spin asymmetry (SSA) for di-hadron production as follows:
\begin{eqnarray}
A_{UT}(\phi_{kS},k_{T} ) & = & \frac{\dsigminus }{\dsigsum } \\ \nonumber
 & \propto & \frac{\Delta^{N}f_{g/p^{\uparrow}}(x,\kp)}{2f_{g/p}(x, \kp)},
\label{eqn:aut}
\end{eqnarray}
where subscript ``U" represents the unpolarized electron beam and ``T" indicates
the transverse polarization of the proton beam.
$\phi_{kS}=\phi_{k_{T}}-\phi_{S}$ stands for the angular difference between the
total di-hadron transverse momentum $\vec{k}_{T}$ and the polarized proton spin
direction $\vec{S}_{\perp}$. The amplitude of the SSA is proportional to the
corresponding Sivers function divided by the unpolarized parton distributions.

\section{Monte Carlo simulation setup}
\label{sec:mcsetup}
\begin{figure}
\begin{center}
\includegraphics[width=0.45\textwidth]{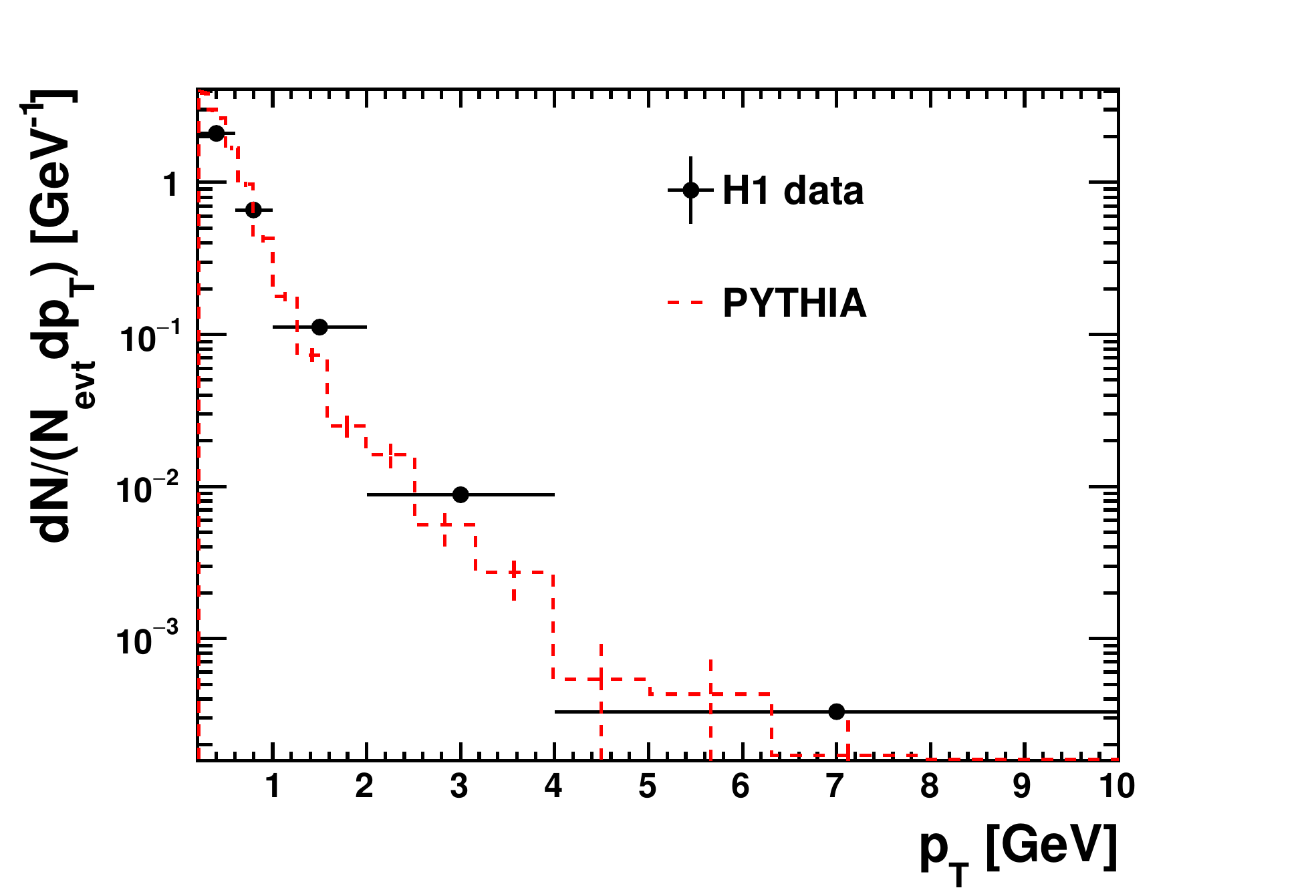}
\end{center}
\caption{(Color online) Charged particle transverse momentum distributions for
	$0<\eta<1.5$ defined in the virtual photon-hadron center of mass frame. The
	HERA data~\cite{Alexa:2013vkv} for 5 GeV$^{2}<Q^{2}<10$ GeV$^{2}$,
	$0.0005<x_{B}<0.002$ with a beam energy 27.6 GeV $\times$ 920 GeV are compared
	to the tuned PYTHIA results.}
\label{fig:hera_tuning_pt}
\end{figure}

\begin{figure}
	\begin{center}
		\includegraphics[width=0.45\textwidth]{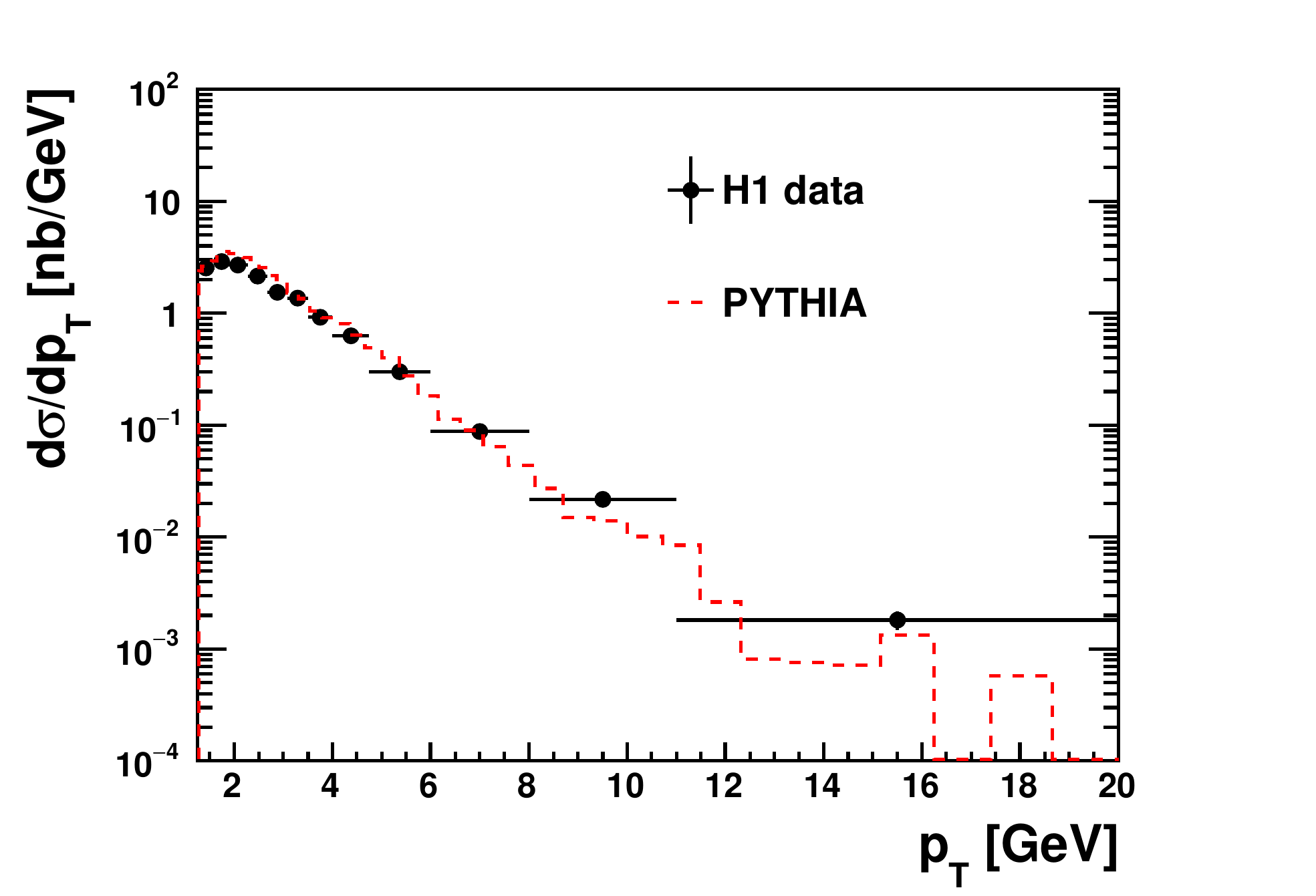}
	\end{center}
\caption{(Color online) $D^*$ transverse momentum distributions for $|\eta_{Lab}|<1.8$ defined 
	in the virtual photon-hadron center of mass frame. The HERA data~\cite{Aaron:2011gp} for
	5 GeV$^{2}<Q^{2}<100$ GeV$^{2}$, $0.02<y<0.7$ with the beam energy 27.6 GeV $\times$ 920 GeV
	are compared to the tuned PYTHIA results.}
	\label{fig:hera_tuning_pt_Dstar}
\end{figure}

In this section, we will describe the setup for our event generation. We use the
PYTHIA-6.4 Monte Carlo (MC) program~\cite{Sjostrand:2006za} to simulate the
unpolarized cross section as expected at an EIC. The PYTHIA generator has been
found to reproduce the charged and open charm particle production in the
electron proton collisions at HERA. The comparison of the HERA
data~\cite{Alexa:2013vkv,Aaron:2011gp} and the output of the tuned PYTHIA MC for
charged particles and $D^*$ mesons is shown in Fig.~\ref{fig:hera_tuning_pt} and
Fig.~\ref{fig:hera_tuning_pt_Dstar}. Based on this reasonable description of the
unpolarized DIS cross section, we will discuss our strategy to obtain the SSA
based on weighting the unpolarized results from PYTHIA.
 
\begin{figure}
	\begin{center}
		\includegraphics[width=0.45\textwidth]{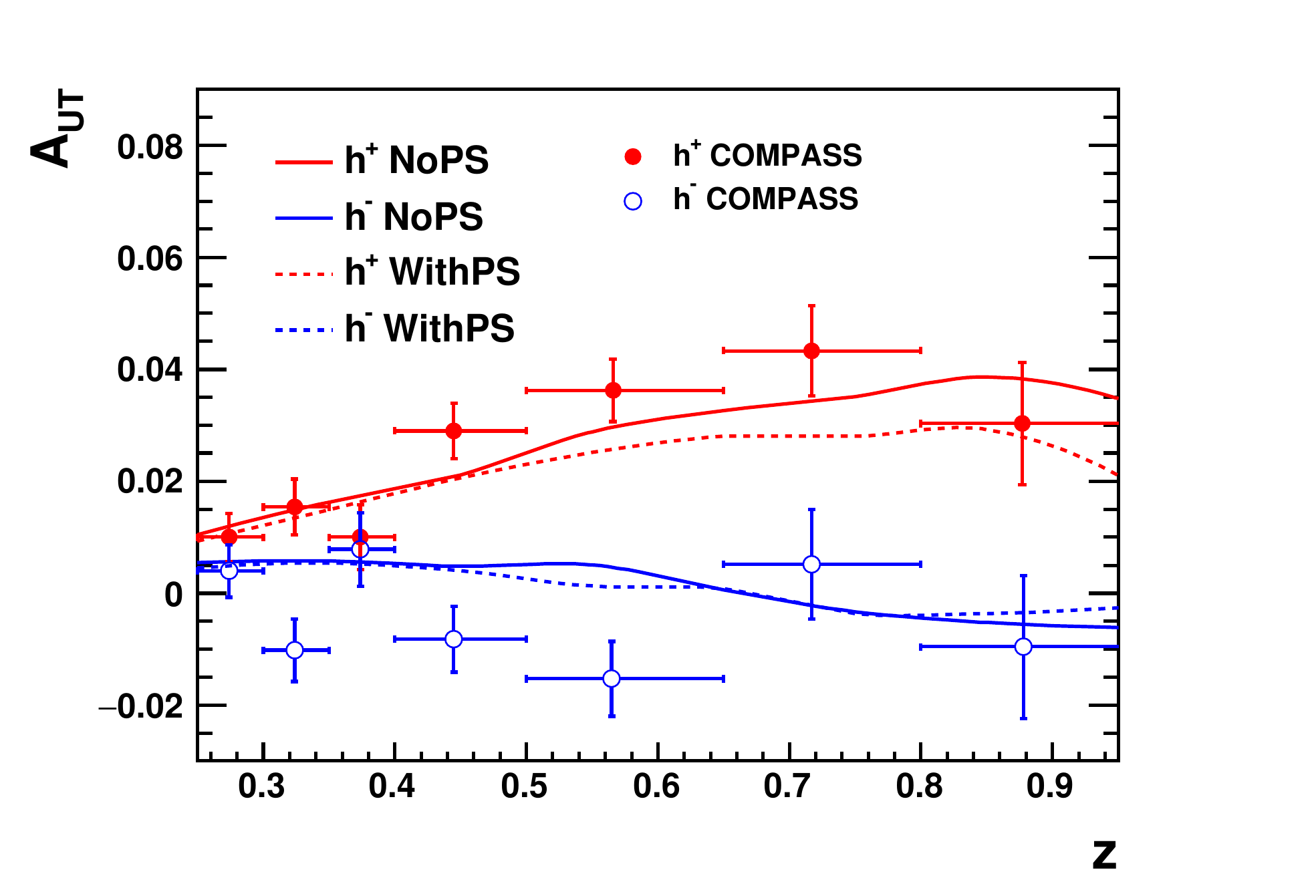}
	\end{center}
\caption{(Color online) Comparison of the charged hadron asymmetry measured by COMPASS
 with the one from the weighting method. The COMPASS data are taken from Ref.~\cite{Adolph:2012sp}.
 Radiation effects are estimated by turning on (WithPS) and off (NoPS) parton shower mechanism in   
 PYTHIA shown with dotted and solid lines in this comparison.}
	\label{fig:compass}
\end{figure}

In the simulation, we model the amplitude of the asymmetry as an incoherent superposition of all 
contributing subprocess on the event-by-event basis. For every event, a weighting factor is 
obtained according to the kinematics and parton flavor as follows:
\begin{equation}
w=\frac{\Delta^{N} f_{a/p^{\uparrow}}(x,\kp,Q^{2})}{2f_{a/p}(x,\kp,Q^{2})}.
\end{equation} 
At the end, the Monte Carlo asymmetry can be understood as the weighted sum of the asymmetry 
weights from signal (gluon initiated channels) and background (quark initiated channels) processes similar to the strategy used in Ref.~\cite{Airapetian:2010ac}: 
\begin{equation}
A_{UT}=R_{g}\frac{\Sigma_{i}^{N_g} w_i}{N_g}+R_{q}\frac{\Sigma_{i}^{N_q} w_i}{N_q},
\end{equation}
in which $N_g$ and $N_q$ indicate the number of gluon and quark initiated events in
the analyzed event sample. The corresponding event fraction is thus obtained as
$R_g = N_g /(N_g+N_q)$ and $R_q = N_q/(N_g+N_q)$.
In the experiment, it is very hard to reliably separate different subprocesses. Therefore, 
the fractions of events from different subprocesses are modeled using PYTHIA in this analysis. A validation of this weighting method against experimental data from COMPASS~\cite{Adolph:2012sp} is discussed at the end of this section (see Fig.~\ref{fig:compass}). 

The parameterization of the Sivers function is given in a factorized form as
\begin{eqnarray}
& & \Delta f_{a/p^{\uparrow}}=2\mathcal{N}_{a}(x_a)f_{a/p}(x_{a}, Q^{2})h(k_{\perp})\gauskp, \\
& & \mathcal{N}_{a}(x_a)=N_a x^{\alpha_a}(1-x)^{\beta_a}\frac{(\alpha_a+\beta_a)^{(\alpha_a+\beta_a)}}{\alpha_a^{\alpha_a}\beta_a^{\beta_a}}, \\
& & h(\kp)=\sqrt{2e}\frac{\kp}{M_1}e^{-\kp^2/M_1^2},
\end{eqnarray}
in which $f_{a/p}(x_{a}, Q^{2})$ is the unpolarized parton distribution, $\mathcal{N}_{a}(x_a)$
and $h(k_{\perp})\gauskp$ describe the $x$ and $k_{\perp}$ dependence of the Sivers function.
The magnitude of the asymmetry from background contributions is calculated from the quark Sivers 
function from these recent fits~\cite{Anselmino:2016uie} listed as:
\begin{eqnarray}
& & N_{u_v}=0.18, \quad \alpha_{u_v}=1.0, \quad \beta_{u_v}=6.2, \\ \nonumber
& & N_{d_v}=-0.52, \quad \alpha_{d_v}=1.9, \quad \beta_{u_v}=10.0, \\ \nonumber
& & N_{\bar{u}}=-0.01, \quad  N_{\bar{d}}=-0.06, \quad M_1^2 = 0.8 \ \mathrm{GeV}^2
\end{eqnarray}

For the gluon Sivers function we utilize two models as input to our study. The first model 
is the SIDIS1 set obtained in the fit~\cite{D'Alesio:2015uta}, which follows a similar
parameterization form as the quark Sivers function with the parameters given by:
\begin{equation}
N_g = 0.65, \alpha_g=2.8, \beta_g=2.8, M_g^2=0.43 \ \mathrm{GeV}^{2} 
\end{equation}

\begin{figure}
	\centering
	\subfigure[]{
		\includegraphics[width=0.45\textwidth]{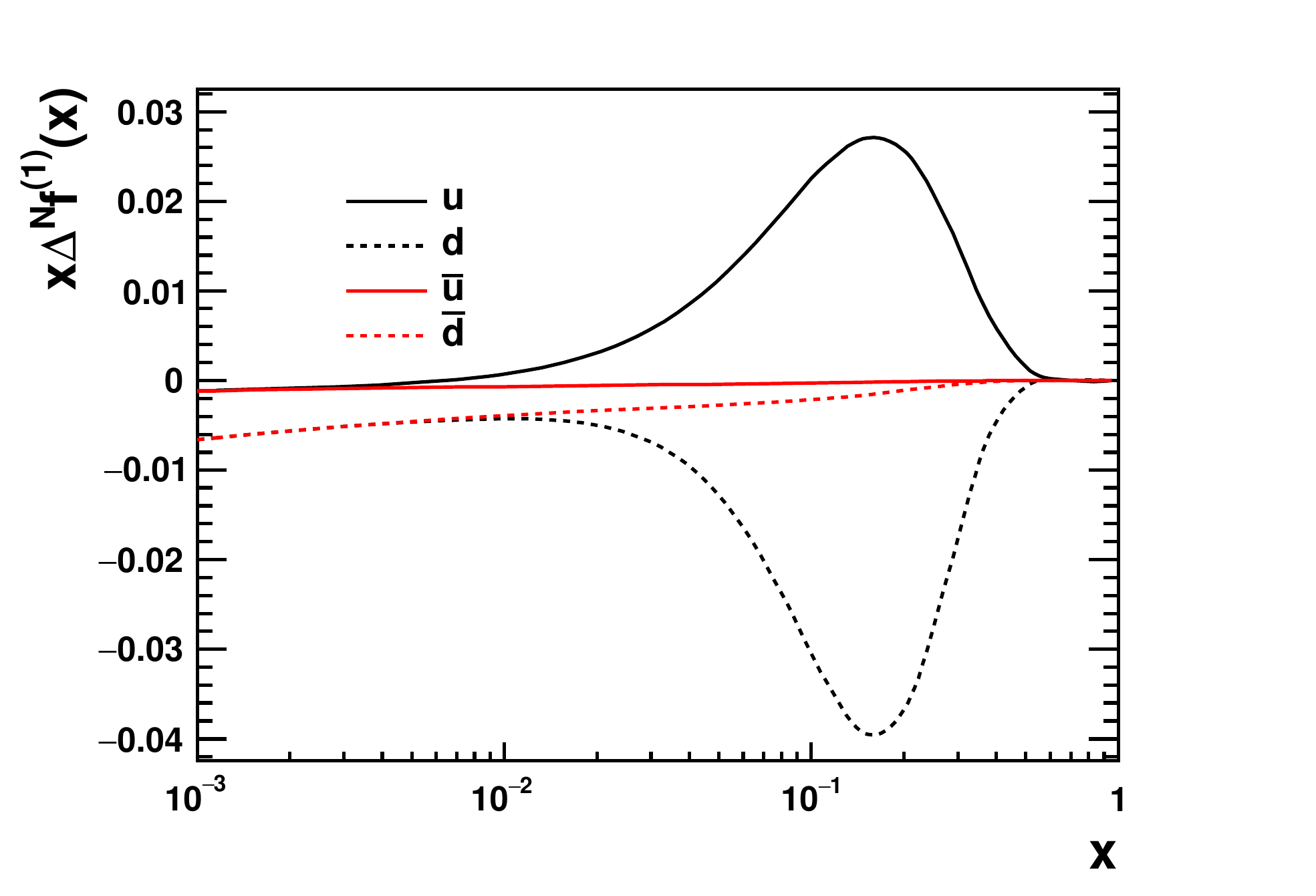}
		\label{fig:quark_sivers_input}
	}
	\subfigure[]{
		\includegraphics[width=0.45\textwidth]{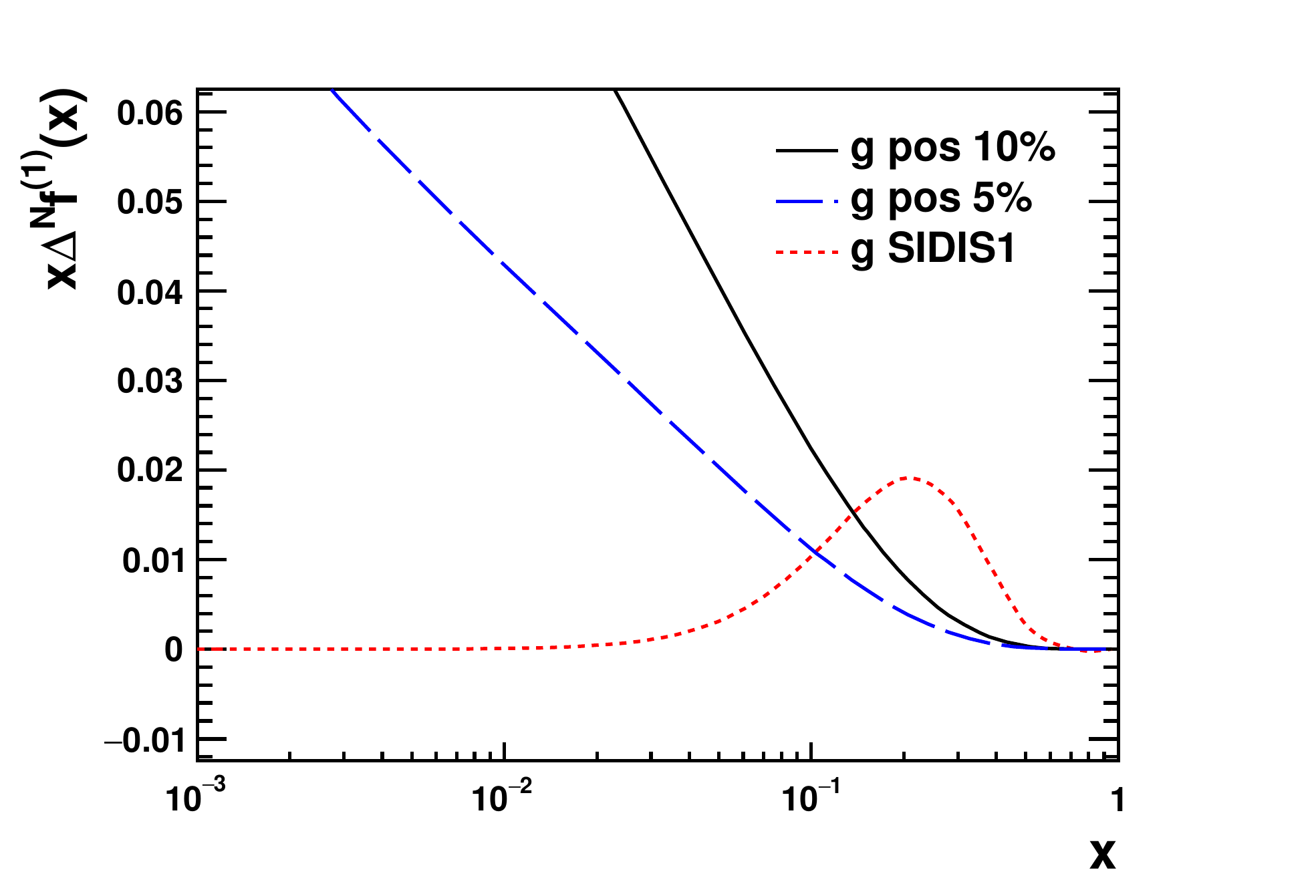}
		\label{fig:gluon_sivers_input}
	}
\caption{(Color online) The first $\kp$ moment $x\Delta^N f^{(1)}(x)$ of Sivers function used in this work for quarks (a) and gluon (b) varying with $x$ at the scale $Q^2=4$ GeV$^2$. Sivers moments for $u$, $d$, $\bar{u}$ and $\bar{d}$ are displayed by the black solid, dotted and red solid, dotted lines in (a). Solid, dashed and dotted line in (b) represents the gluon Sivers on positivity bound by 10\%, 5\% and the SIDIS fit from Ref.~\cite{D'Alesio:2015uta}. }
	\label{fig:sivers_input}
\end{figure}

The second gluon Sivers model relies on the positivity bound assumption used in~\cite{Anselmino:2004nk}: 
\begin{equation} 
f_{1T}^{\perp g}=-\frac{2\sigma M_{p}}{\kp^{2}+\sigma^{2}}f_{g}(x,\kp), \quad \sigma=0.8 \ \mathrm{GeV},
\end{equation} 
in which the positivity limit is saturated when $\kp=0.8$ GeV. We will use 10\%
and 5\% of the positivity bound to study quantitatively the measurability of the
gluon Sivers function. We calculate the weight of every event according to the
inputs discussed here to obtain the magnitude of the asymmetry in the final
state. An example of the first $\kp$ moment of the input Sivers distribution
$\Delta^N f^{(1)}(x)=\int d^2\kp\frac{\kp}{4m_p}\Delta^N
f_{a/p^{\uparrow}}(x,\kp)$ is shown in Fig.~\ref{fig:sivers_input}.
Fig.~\ref{fig:quark_sivers_input} shows the quark Sivers functions used to
estimate the background contribution, while the gluon Sivers functions are shown
in Fig.~\ref{fig:gluon_sivers_input}. For the current parameterizations the
quark Sivers functions are maximum for $x>$0.1 for the valence quarks and become
negligible in the small $x$ regime. The magnitude of the sea-quark sivers
functions is small over the entire $x$-range. It is noticeable that the gluon
Sivers functions based on the positivity bound assumption and SIDIS1 set have
quite a different functional form in $x$.

We provide in Fig.~\ref{fig:compass} a comparison of the charged hadron
asymmetry measured by the COMPASS experiment~\cite{Adolph:2012sp} with the
asymmetry obtained from weighting PYTHIA events according to the method
described above with the quark Sivers functions. It is not surprising to see
that radiation effects modeled by the parton shower mechanism in PYTHIA are
quite weak at the COMPASS energy. The comparison also shows that one can 
describe both positive and negative charged hadron asymmetries
from COMPASS with the event weighting method.

It should be explicitly noted that the parameterizations of the Sivers asymmetry
discussed here are not accounting for any effects due to the QCD scale dependence of TMDs. 
The QCD evolution of the Sivers function can be calculated in the QCD resummation formalism following the Collins-Soper-Sterman method~\cite{Collins:2014jpa,Collins:1984kg} by applying
the correct Sudakov factor to the spin dependent parton distributions~\cite{Kang:2011mr, Echevarria:2014xaa, Anselmino:2012aa}. 
The needed precise phenomenological inputs to determine the QCD scale evolution of Sivers 
asymmetry is not yet available, while they can be obtained from the future RHIC and EIC measurements. We will therefore not address the evolution of the gluon Sivers function in this paper but leave it for future work.

In order to estimate the statistical uncertainty of the SSA in our simulation,
we use $\delta A=\sqrt{\frac{1}{P^{2}N}-\frac{A^2}{N}}$ from Ref.~\cite{Bunce:2000uv},
where $N$ represents the count of selected pairs in a certain kinematic bin, and
$P$ indicates the polarization of the proton beam. In this work, we assume a polarization 
$P=$ 70\% for the EIC beam energy configuration of 20 GeV $\times$ 250 GeV with an integrated luminosity $\mathcal{L}_{int}=10$ fb$^{-1}$. 

\section{Results and Discussions}
\label{sec:results}

In this study, the event kinematics has been restricted to $0.01<y<0.95$ and
$1\, \textrm{GeV}^{2}<Q^{2}<20 \, \textrm{GeV}^{2}$ with the electron and proton
beam energy configuration of 20 GeV $\times$ 250 GeV. A detector system especially
designed for EIC with a wide acceptance $-4.5<\eta_{Lab}<4.5$ for measuring charged
particles~\cite{Aschenauer:2014cki} has been assumed, in which case the event
kinematics can be well reconstructed from the scattered electron. This selection
gives an average event kinematics for the minimum bias events as 
$\langle x_{B} \rangle = 0.0012$, $\langle Q^{2} \rangle = 2.5$ GeV$^{2}$, 
$\langle W \rangle = 54.6$ GeV. The wide kinematic reach at this high center-of-mass 
energy ($\sqrt{s}$=141 GeV) makes it possible to study the evolution of the Sivers asymmetry 
and to access the region dominated by gluons.

As discussed in Sec.~\ref{sec:gluon_sivers}, there is a correlation between the
vector sum of the transverse momentum $k_{T}$ for the selected hadron pairs or
di-jets with the initial transverse motion of gluons. It then follows naturally
to investigate the Sivers asymmetry through the sine-modulation for the angle
$\phi_{kS} = \phi_{k_{T}}-\phi_{S}$, which defines the difference between $k_T$
and the spin direction of the proton. To tag the gluon distributions and study the
Sivers asymmetries we will study the $D$ meson pair, charged di-hadrons and
di-jet production. The detailed experimental cuts for each channel are listed as follows:
\begin{itemize}
	 \item $D$ meson pair production\\
	  $D^{0}\rightarrow \pi K $, $|\eta^{\pi/K}_{Lab}|<3.5$, $p_{T Lab}^{\pi/K}>0.2$ GeV,               $p^{D}_{T}>0.7$ GeV, $z_{D}>0.1$ 
	 \item charged di-hadron production\\
	  $-4.5<\eta_{Lab}<4.5$, $p^{h}_{T}>1.4$ GeV, $z_{h}>0.1$ 
	 \item di-jet production\\
	  $\pi,K,p,\gamma$ with $p_{T Lab}>0.25$ GeV, $|\eta_{Lab}|<4.5$ used for the jet 
      reconstruction with anti-$k_{T}$ algorithm and a cone radius $R=1$ ,the trigger jet has 
      $p_{T}^{jet1}>4.5$ GeV and the associate jet $p_{T}^{jet2}>4$ GeV
\end{itemize}

\subsection{The Gluon SSA in Open Charm Production}
\label{subsec:dicharm}

The heavy flavor production has been proven to be very useful for measuring the
gluon Sivers asymmetry in proton-proton collisions as shown in
Ref.~\cite{Anselmino:2004nk,DAlesio:2017rzj}. Similar to the case in
hadron-hadron reactions, it is well accepted that the heavy flavor production in
DIS is a very clean channel to directly probe the gluon distributions. In this
section we demonstrate the possibility to measure the gluon Sivers function in
open charm production $\gamma^{*} g\rightarrow c \bar{c}$ with $D^{0}$-mesons in
the final state. Open charm production has the advantage that quark initiated
process are strongly suppressed and one becomes essentially only sensitive to
gluon initiated subprocesses. The $D^{0}$-mesons are identified through the $\pi
K$ decay channel by taking advantage of the vertex tracking detector integrated to the main detector.
\begin{figure}[htb]
	\centering
	\subfigure[]{
	 \includegraphics[width=0.45\textwidth]{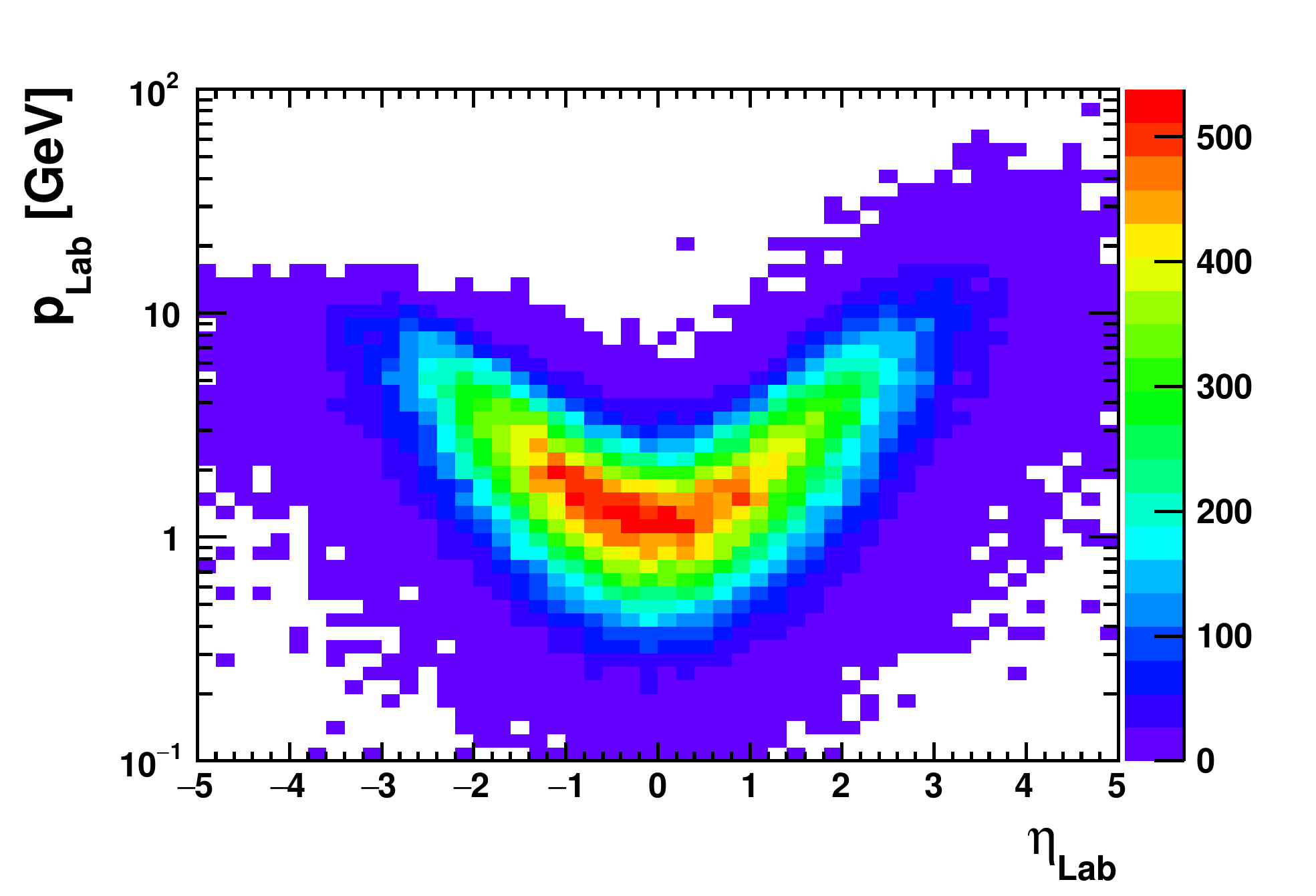}
	}
\caption{(Color online) Momentum-pseudorapidity correlation for $K$ decayed from $D^{0}$ meson.
	}
	\label{fig:d_decay_kaon}
\end{figure}
The momentum and pseudorapidity distribution of the $K$-meson from the $D^{0}$ decay can
be found in Fig.~\ref{fig:d_decay_kaon}. The $K$ momenta are typically a few GeV
in the central rapidity region, and extend to 10 GeV at rapidities
$|\eta|>1$. The distribution for the $\pi$-mesons from the $D^0$ decay is found to be 
very similar to the one from $K$s.
\begin{figure}[htb]
  \centering
  \includegraphics[width=0.45\textwidth]{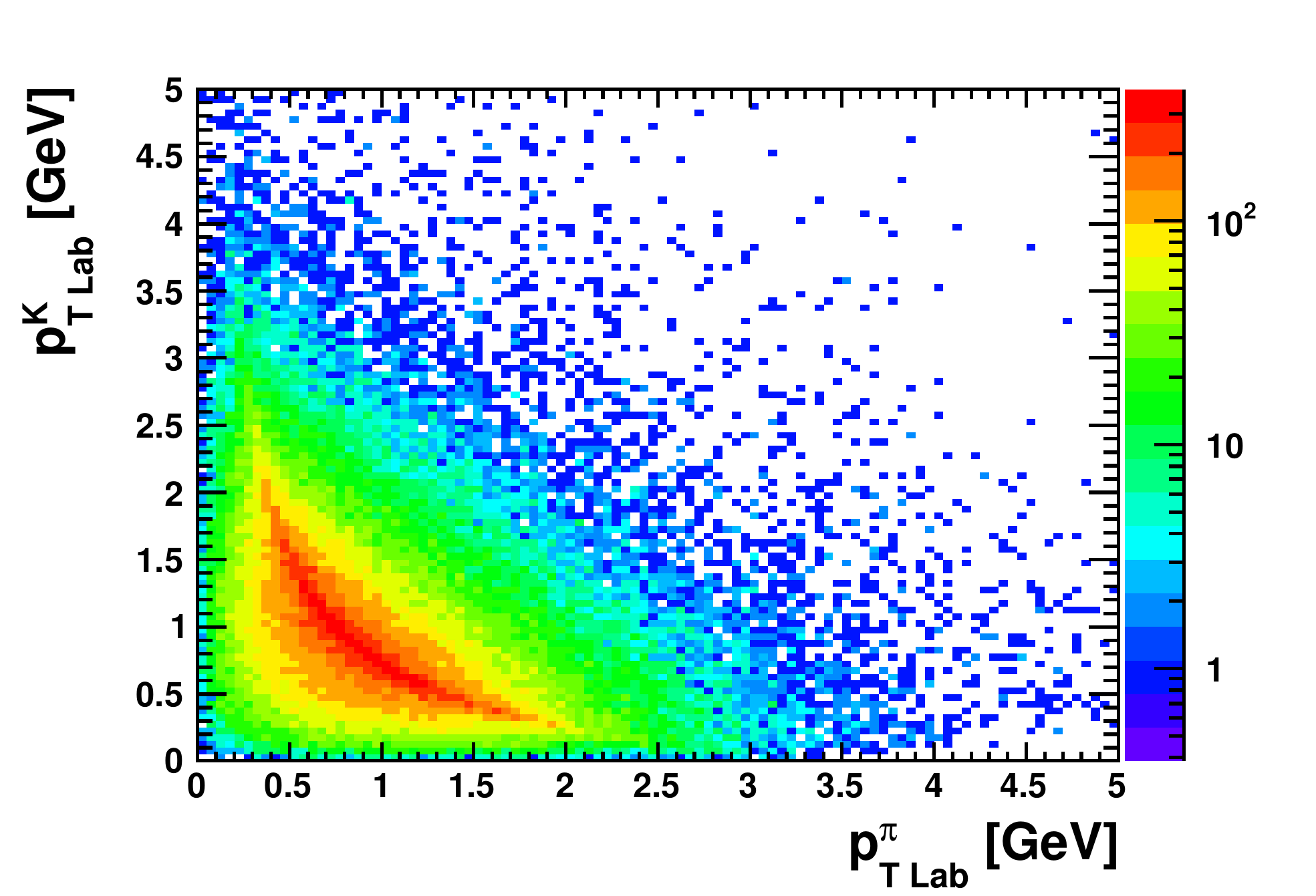}
  \caption{(Color online) $p_{T}$ correlation for $\pi$ and $K$ decayed from the same $D$ meson.
  }
  \label{fig:d_decay_pionkaon_pt}
\end{figure}
The $D^{0}$ meson decay products are required to be in $|\eta^{\pi/K}_{Lab}|<3.5$ and to have transverse momenta $p^{\pi/K}_{TLab}>0.2$ GeV to be reconstructed and identified.
The $p_{T Lab}$ correlation of the $D^{0}$ meson decay products is shown in
Fig.~\ref{fig:d_decay_pionkaon_pt}, most of the $\pi$, $K$ products pass the
transverse momentum cut. The kinematics of a directly produced $D^{0}$-meson and
for one from the decay of a heavier charm-mesons is basically the same, therefore all
$D^{0}$-mesons with $p_T>0.7$ GeV and $z_h>0.1$ are included in this study.

\begin{figure}[htb]
  \centering
  \includegraphics[width=0.45\textwidth]{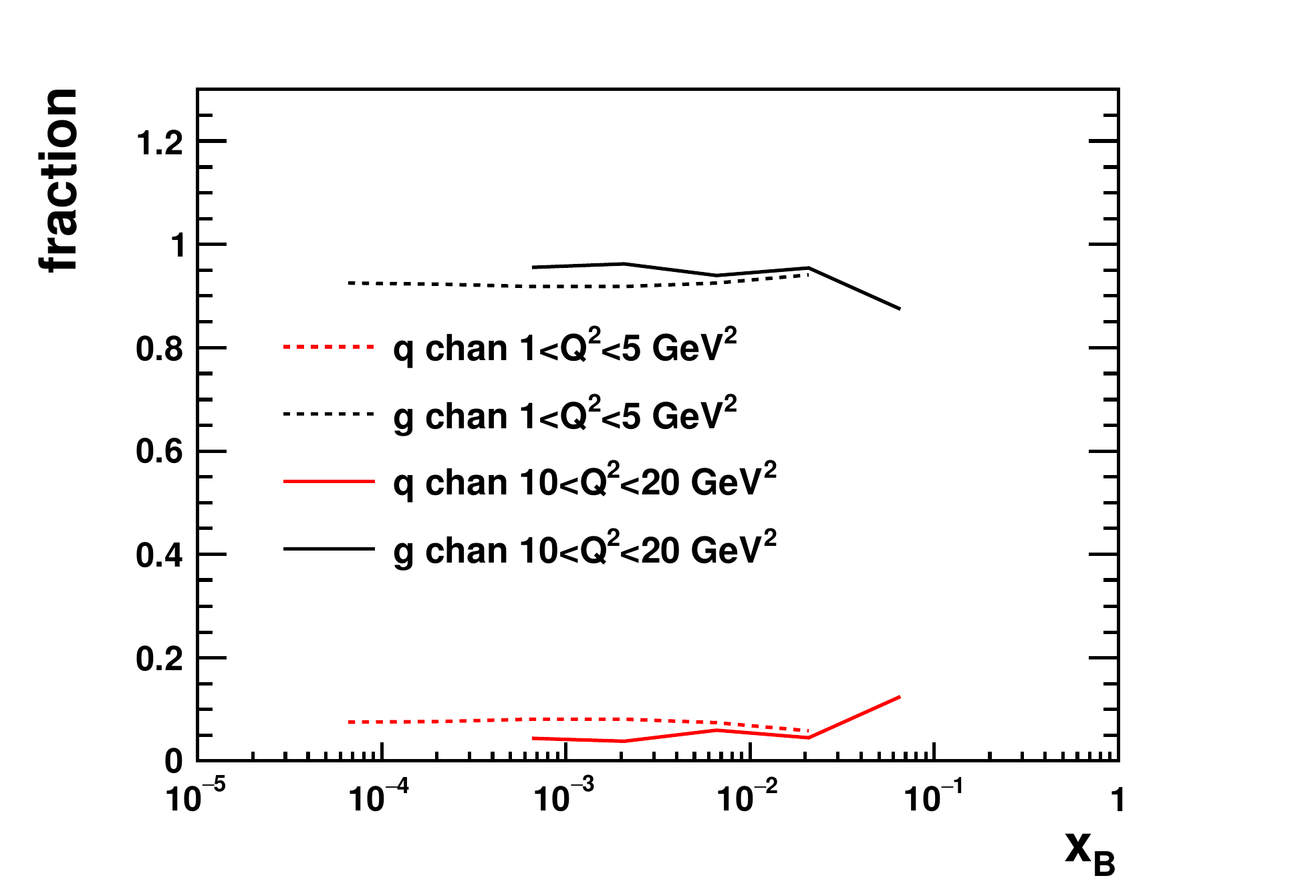}
  \caption{(Color online) Fraction of underlying subprocesses initiated by quarks (red curves)
   or gluons (black curves) for $D\bar{D}$ pair production. The solid and dotted curves represent 
   the two $Q^2$ bins of $10<Q^{2}<20$ GeV$^2$ and $1<Q^{2}<5$ GeV$^2$. }
	\label{fig:d_pair_proc_fraction}
\end{figure}

To capture the full charm anti-charm quark pair kinematics, we select events
with $D\bar{D}$ pairs in the final state. Fig.~\ref{fig:d_pair_proc_fraction}
shows that gluon initiated processes account for about 90\% of the total
selected events over a wide range in $x_B$ for two $Q^2$ bins. For $x_B >$ 0.1
quark initiated subprocesses become slightly more important.

\begin{figure}[ht]
	\centering
	\subfigure[]{
		\includegraphics[width=0.45\textwidth]{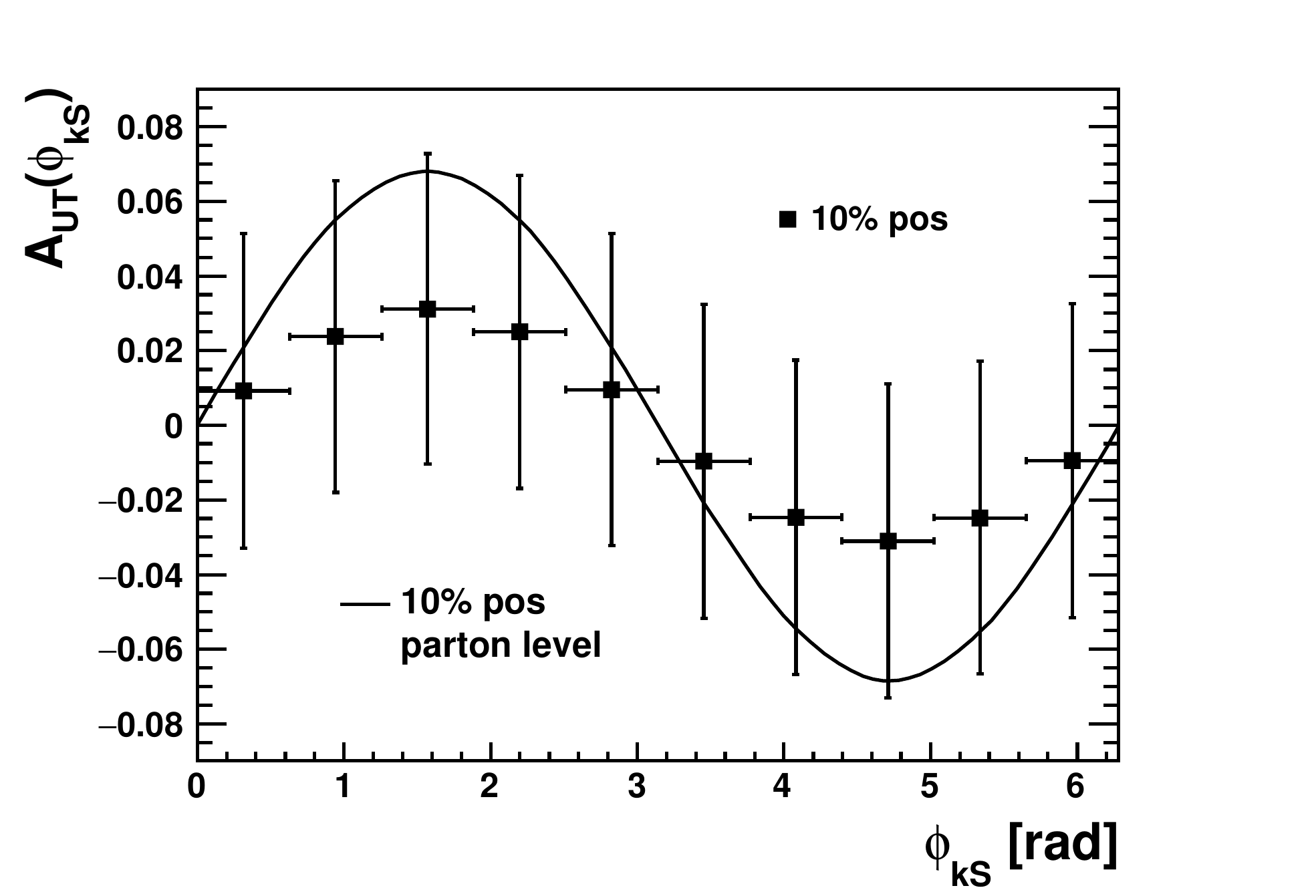}
		\label{fig:ddbar_ssa_phisk}
	}
	\subfigure[]{
		\includegraphics[width=0.45\textwidth]{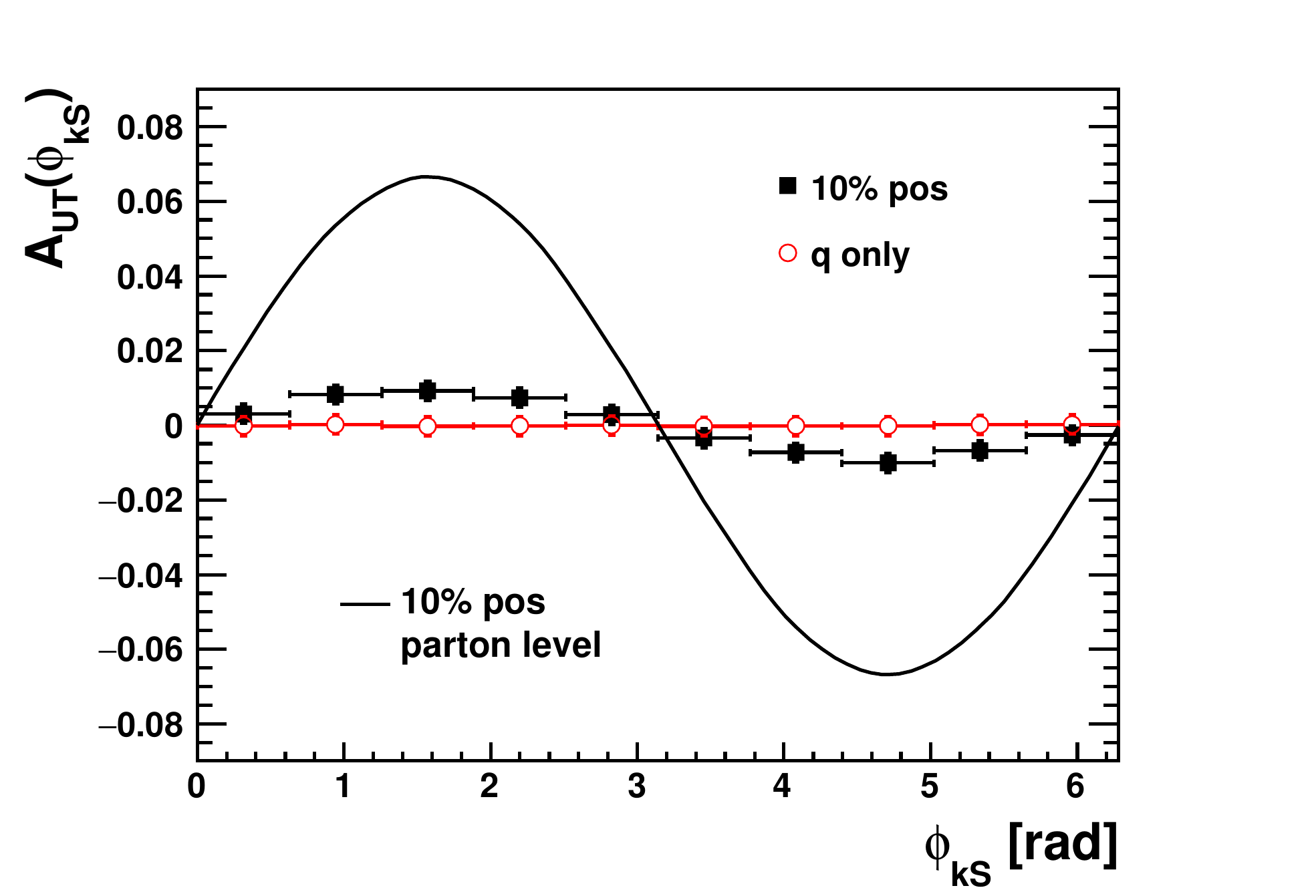}
		\label{fig:dsingle_ssa_phisk}
	}	
	\caption{(Color online) Projection for the SSA dependence on $\phi_{ks}$ 
	  shown for different input for the gluon Sivers function for $D\bar{D}$ pairs (a) and single 
      $D$ mesons (b). The vertical bars represent the statistical uncertainties obtained with the 
      kinematic cuts $|\eta^{\pi/K}_{Lab}|<3.5$, $p^{\pi/K}_{T Lab}>0.2$ GeV, $z^D>0.1$, 
      $p_{T}^D>0.7$ GeV, $0.01<y<0.95$ and $1\, \textrm{GeV}^{2}<Q^{2}<20 \, \textrm{GeV}^{2}$ 
      at the electron-proton beam energy 20 GeV$\times$ 250 GeV and an integrated luminosity       
      $\mathcal{L}_{int}=10$ fb$^{-1}$. The initial parton level asymmetry is shown with the solid 
      line. The Sivers asymmetry for 10\% of the positivity bound and for the quark contribution 
      are displayed with the closed squares and the open circles, respectively.		
	}
\end{figure}

The sensitivity of the $D\bar{D}$ pair measurement to the magnitude of the gluon
Sivers function is shown in Fig.~\ref{fig:ddbar_ssa_phisk}. The statistical
uncertainty is based on an integrated luminosity of $\mathcal{L}_{int}=10$ fb$^{-1}$. 
The solid curve represents the parton level asymmetry. The Sivers asymmetry based on the 
scenario with 10\% of the positivity bound of the gluon Sivers function is indicated by the 
black filled symbols. The SSA for the background quark initiated Sivers effect is consistent 
with zero and not shown here. The limited statistical precision for the $D\bar{D}$ final 
state due to the small branching ratio (3.87\% $D\rightarrow K\pi$) makes it challenging to precisely determine the gluon Sivers function on the level of 10\% of the positivity
bound. Therefore, we also investigated the sensitivity to the magnitude of the
gluon Sivers function requiring only one $D$ meson. The Sivers angle $\phi_{kS}$
is calculated replacing $k_T$ with the $D$ meson transverse momentum.
Fig.~\ref{fig:dsingle_ssa_phisk} depicts the SSA based on the 10\% of the gluon Sivers positivity bound assumption, which can be well distinguished from the background SSA due to quark Sivers effects. Comparing Fig.~\ref{fig:dsingle_ssa_phisk} and Fig.~\ref{fig:ddbar_ssa_phisk}, one can
observe that the initial parton level asymmetries are similar, but the magnitude of the final state asymmetry for single $D$ mesons is reduced since the transverse momentum of one $D$ meson is not a good proxy for the initial gluon transverse momentum.

A similar approach to the open charm production is to select $K^+K^-$ pairs in
the final state, which enhances as underlying process $\gamma^* g\rightarrow s\bar{s}$. 
We find in our study this measurement is also statistically limited and can only resolve a gluon Sivers signal up to 10\% of the positivity bound. Since the global features of this measurement 
are similar to the di-hadron case, which will be discussed in the next section, we will not 
provide more information.

\subsection{Gluon SSA through Charged Di-hadron Pairs}
\label{subsec:dihadron}

In SIDIS production, the Leading Order DIS (LODIS) process $\gamma^{*} q\rightarrow q$ is accounting for a large fraction of the charged particle productions. The LODIS process can be largely suppressed by requiring a pair of high $p_T$ charged hadrons.

\begin{figure}[h]
	\centering
	\includegraphics[width=0.45\textwidth]{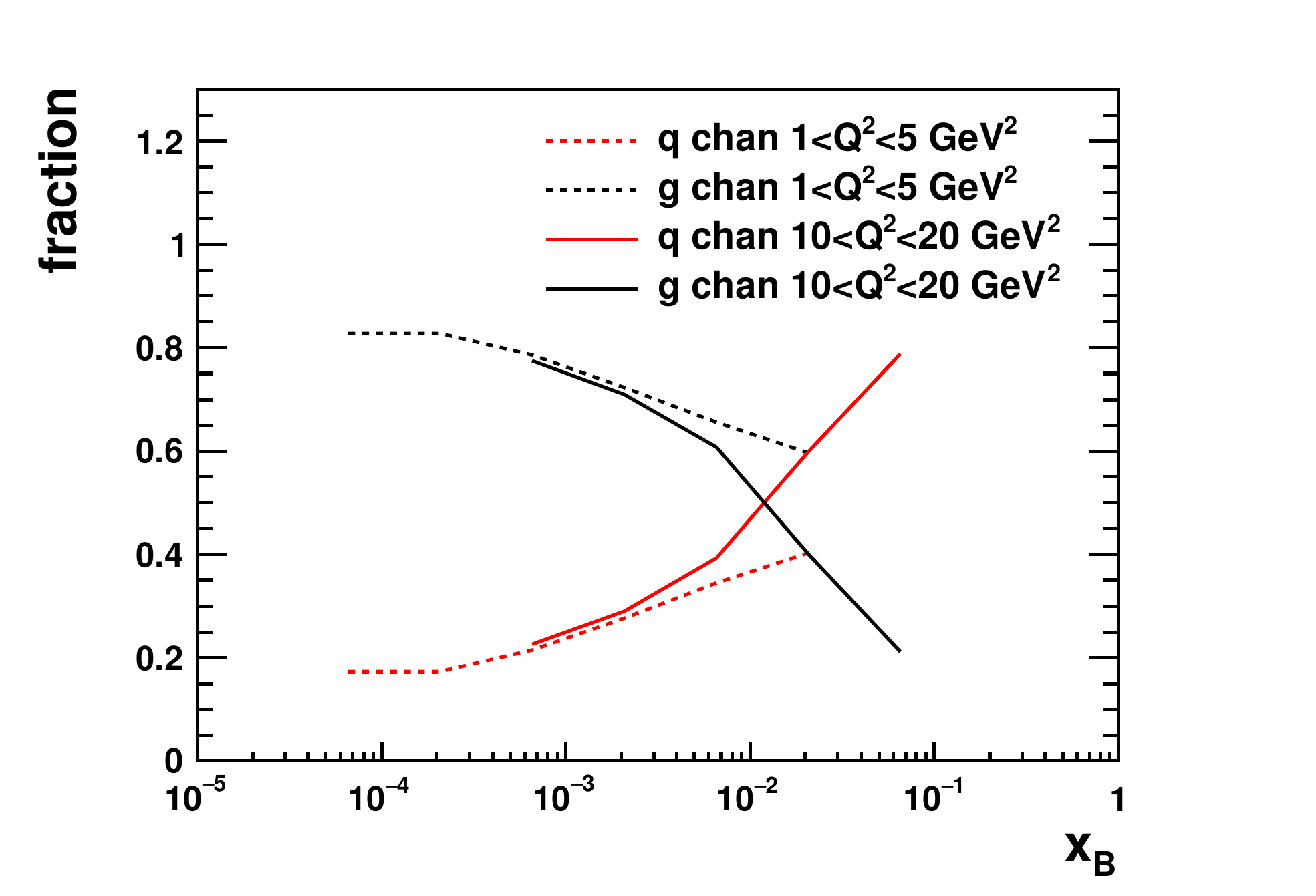}
	\caption{(Color online) Fraction of the underlying subprocesses initiated by quarks 
	(red curves) or gluons (black curves) for high $p_T$ charged di-hadron pairs. 
	The solid and dotted curves represent the $Q^2$ bins of $10<Q^{2}<20$ GeV$^2$
	and $1<Q^{2}<5$ GeV$^2$, respectively }
\label{fig:dihadron_proc_fraction}
\end{figure}

The acceptance for the charged hadrons is required to fit the EIC detector design 
$|\eta_{Lab}|<4.5$ and the hadron pair candidates must have a transverse momentum 
$p^{h}_{T}>1.4$ GeV and $z_{h}>0.1$. To select hadron pairs from
back-to-back jets, we ask $k_{T}<0.7 P_{T}$ with
$P_T=|\vec{p}^{h1}_{T}-\vec{p}^{h2}_{T}|/2$. This way, one can eliminate the
contribution of two hadrons fragmented from the same parton. The event fractions
affected by the gluon and quark initiated processes are shown in
Fig.~\ref{fig:dihadron_proc_fraction}. Around 80\% of the high $p_T$ di-hadron
events are generated from gluon initiated processes in the small $x_B$ region. The
fraction of quark initiated processes grows rapidly as $x_B$ approaches 0.1 and
with increasing $Q^2$. This behavior with $Q^2$ can be understood that more high
$p_T$ hadrons are generated through QCD radiation, which has an increased
probability with increasing $Q^2$.

\begin{figure}
\centering
\subfigure[]{
\includegraphics[width=0.45\textwidth]{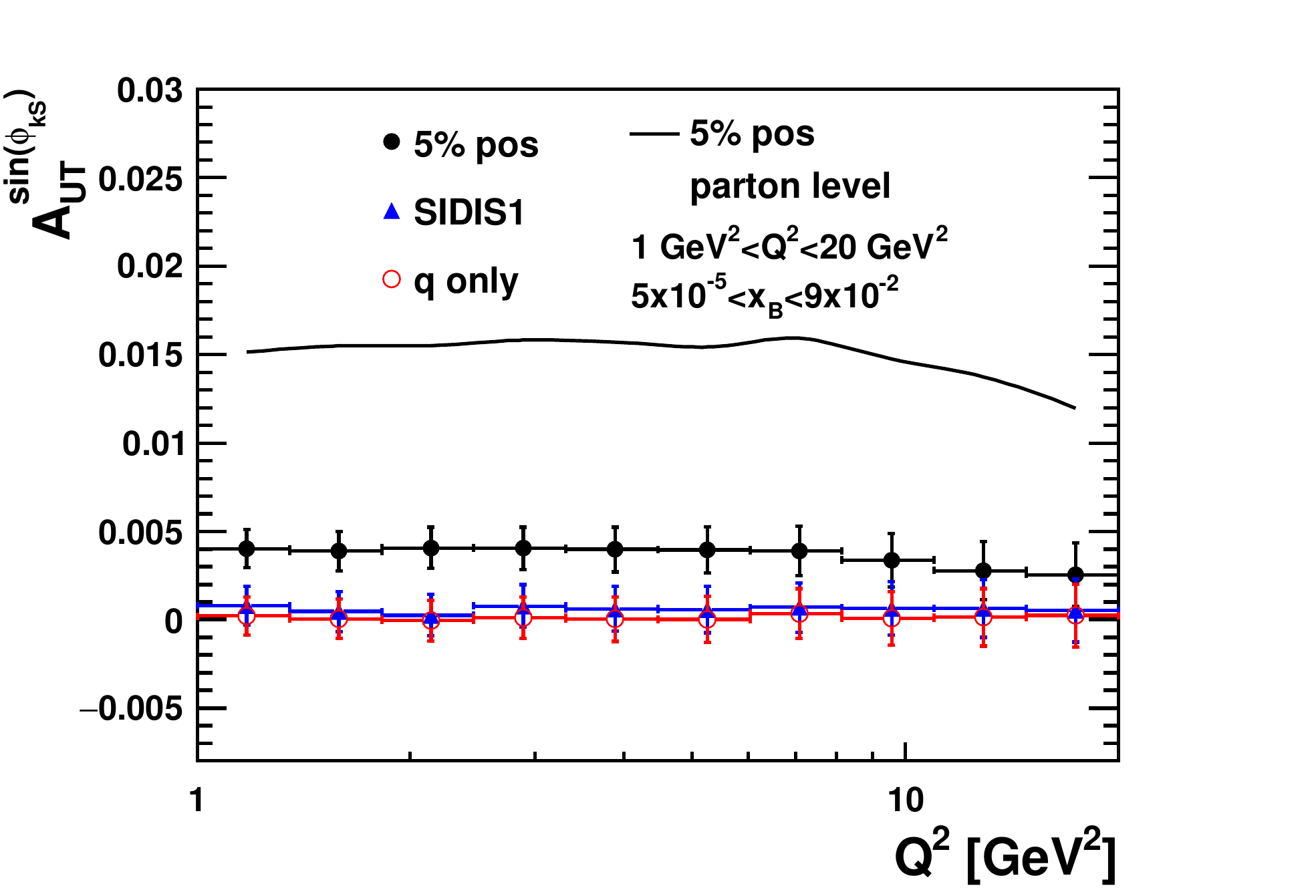}
\label{fig:dihadron_ssq_q2}
}
\subfigure[]{
\includegraphics[width=0.45\textwidth]{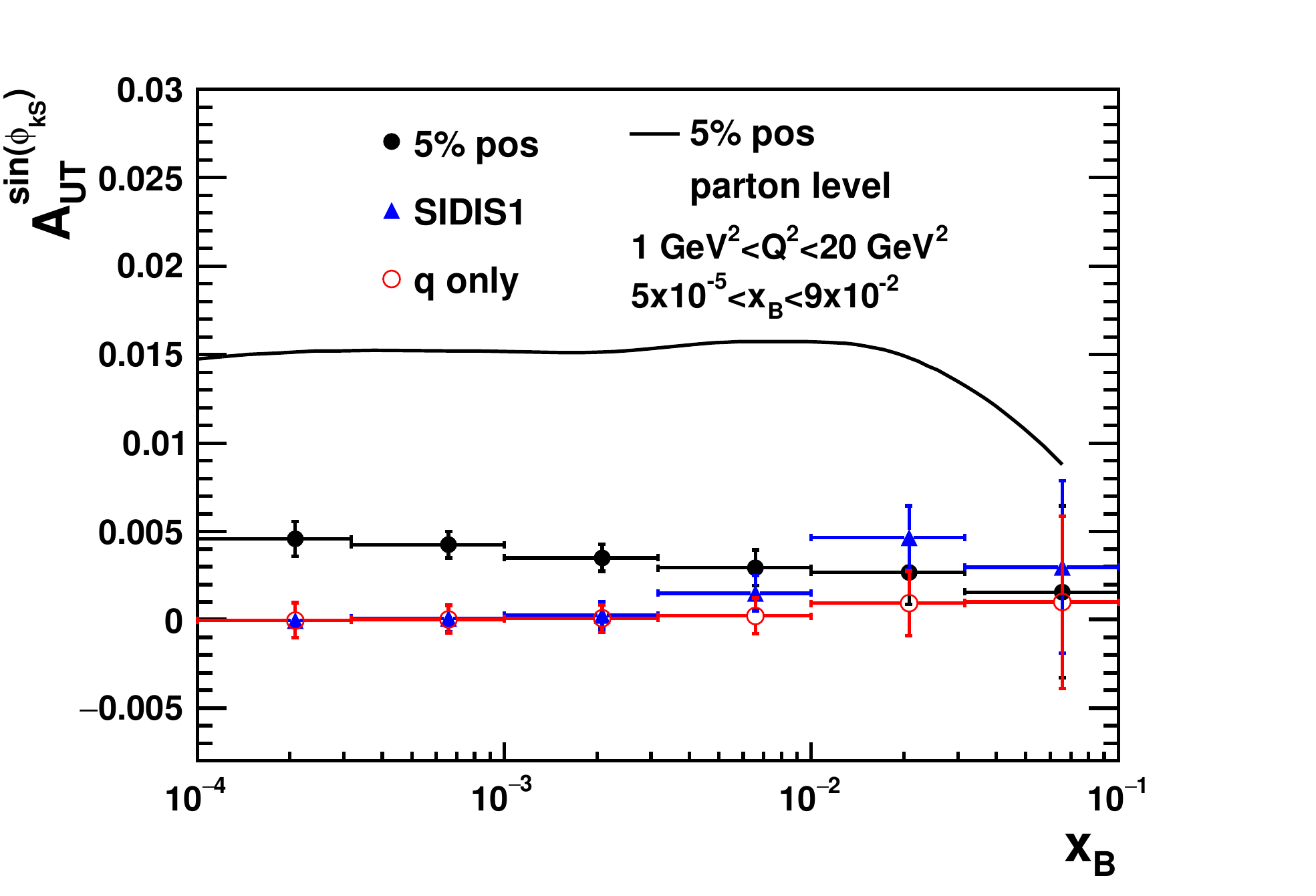}
\label{fig:dihadron_ssa_xbj}
}
\caption{(Color online) Sivers asymmetries dependent on $Q^2$ (a) and $x_{B}$ (b) for charged 
  hadron pairs requiring the cuts $|\eta^{h}_{Lab}|<4.5$, $p^{h}_{T}>1.4$ GeV,
  $z^{h}>0.1$, $k_{T}<0.7P_{T}$, $0.01<y<0.95$ and $1\,\textrm{GeV}^{2}<Q^{2}<20 \,      
  \textrm{GeV}^{2}$ for the electron-proton beam energy 20 GeV$\times$ 250 GeV for an 
  integrated luminosity $\mathcal{L}_{int}=10$ fb$^{-1}$.  }
\label{fig:dihadron_ssa_q2_xbj}
\end{figure}

It is especially noted that an understanding of the gluon Sivers function requires to measure 
its dependence on $x_{B}$ and $Q^2$. Figure~\ref{fig:dihadron_ssa_q2_xbj} compares the SSA 
for charged hadron pairs assuming the magnitude of the gluon Sivers function 5\% of its positivity
bound (solid circles) and the SIDIS1 set (solid triangles) as well as no gluon Sivers contribution, 
but a quark Sivers contribution (open circles). The study shows that a gluon Sivers function of a magnitude of 5\% of the positivity bound can be measured at an EIC. We also find that the initial parton level asymmetry is significantly diluted, a factor 3 as indicated comparing the black curve and
the solid circles. This dilution is larger than the one (factor 2) shown in Fig.~\ref{fig:ddbar_ssa_phisk} for $D\bar{D}$ pairs. The increase can be
explained by the stronger smearing due to the fragmentation of light quarks to hadrons of the correlation between the parton and the hadron $p_T$.

Figure~\ref{fig:dihadron_ssa_q2_xbj} shows the angular modulation of the Sivers function is only 
weakly dependent on $Q^2$, because of the missing TMD evolution in the current framework, 
but much more sensitive on $x_{B}$. The dependence on $x_{B}$ is a natural consequence of the 
behavior of the Sivers function parameterization with $x$.

\subsection{Gluon SSA in Di-jet Production}
\label{subsec:dijet}

Comparing the hadron level observables with jets, it can be clearly seen that
jets provide a more precise reconstruction of the initial gluon kinematics. In
the following we study the sensitivity to the gluon Sivers function in di-jet
production. The jets are reconstructed from charged hadrons ($\pi$, $K$ and
protons) measured in the central tracker together with photons accepted in the
calorimeter requiring a minimum transverse momentum $p_{T Lab}>$ 0.25 GeV and 
$|\eta_{Lab}|<$ 4.5. The jet radius parameter is assumed to be $R=1$ in the
anti-$k_T$ jet reconstruction algorithm. Di-jet events are defined with the trigger
jet $p_T^{jet1}>$ 4.5 GeV and the associate jet $p_T^{jet2}>$ 4 GeV. Similar to
the di-hadron channel, we use the vector sum of the transverse momentum for the
two jets $k_{T}=|\vec{p}_T^{jet1}+\vec{p}_{T}^{jet2}|$ as the proxy to access
the underlying gluon dynamics.
\begin{figure}[h]
	\centering
	\includegraphics[width=0.45\textwidth]{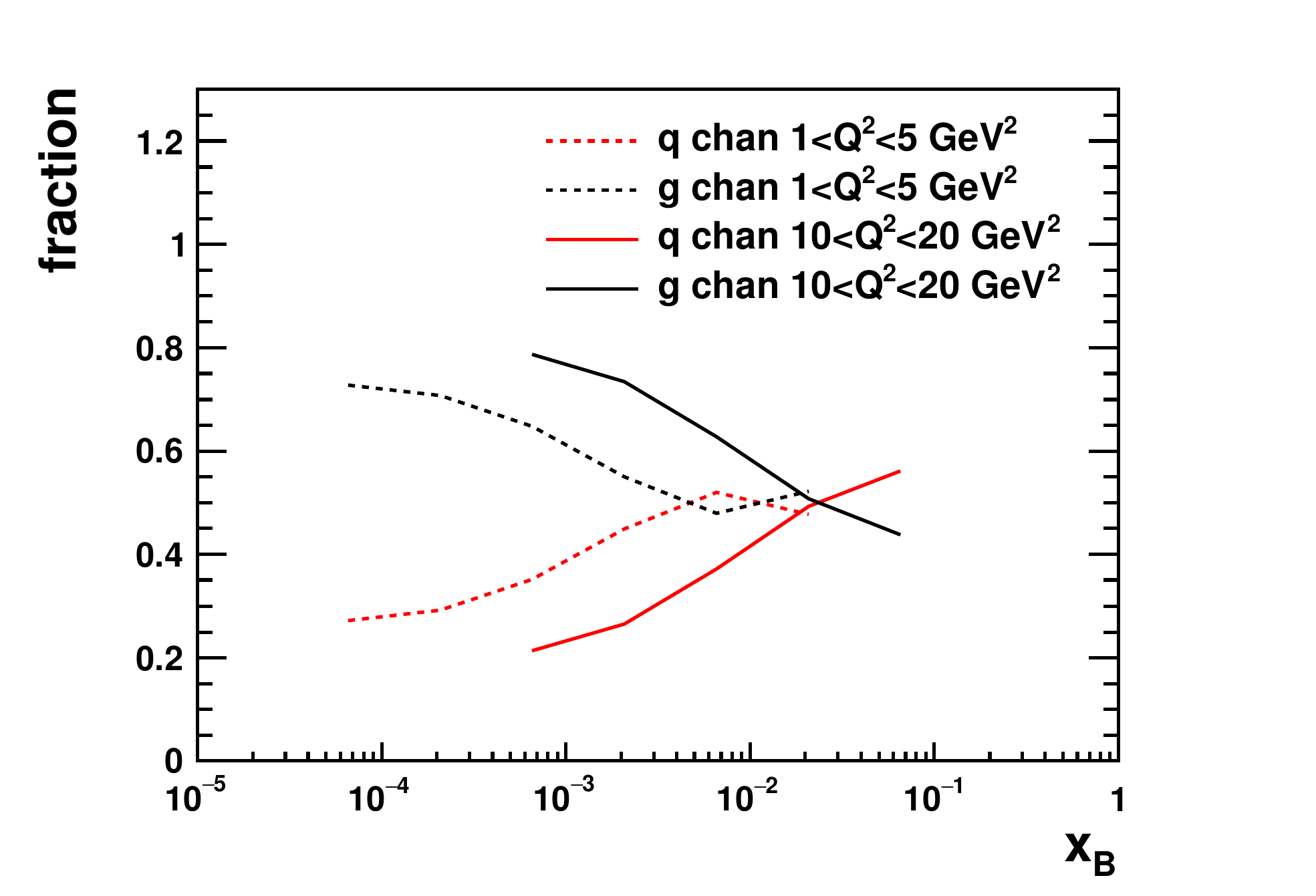}
	\caption{(Color online) Underlying subprocess fraction initiated by quarks (red
	curves) or gluons (black curves) for di-jet production. The solid and dotted
	curves represent the $Q^2$ bins of $10<Q^{2}<20$ GeV$^2$ and $1<Q^{2}<5$
	GeV$^2$, respectively.  } \label{fig:dijet_proc_fraction}
\end{figure}
We present the fractions for quark and gluon initiated processes in Fig.~\ref{fig:dijet_proc_fraction}. The quark fraction is substantial (close to 30\%) for low $Q^2$ events even for $x_{B}\sim 10^{-4}$. The fraction of gluon initiated channels is maximized at small $x_B$ and drops below the quark
fraction if $x_B$ is close to 0.01 or 0.1 depending on the $Q^2$ range. Unlike for the di-hadron 
case, the gluon event fraction increases with $Q^{2}$.

\begin{figure}[h!]
	\centering
	\subfigure[]{
		\includegraphics[width=0.45\textwidth]{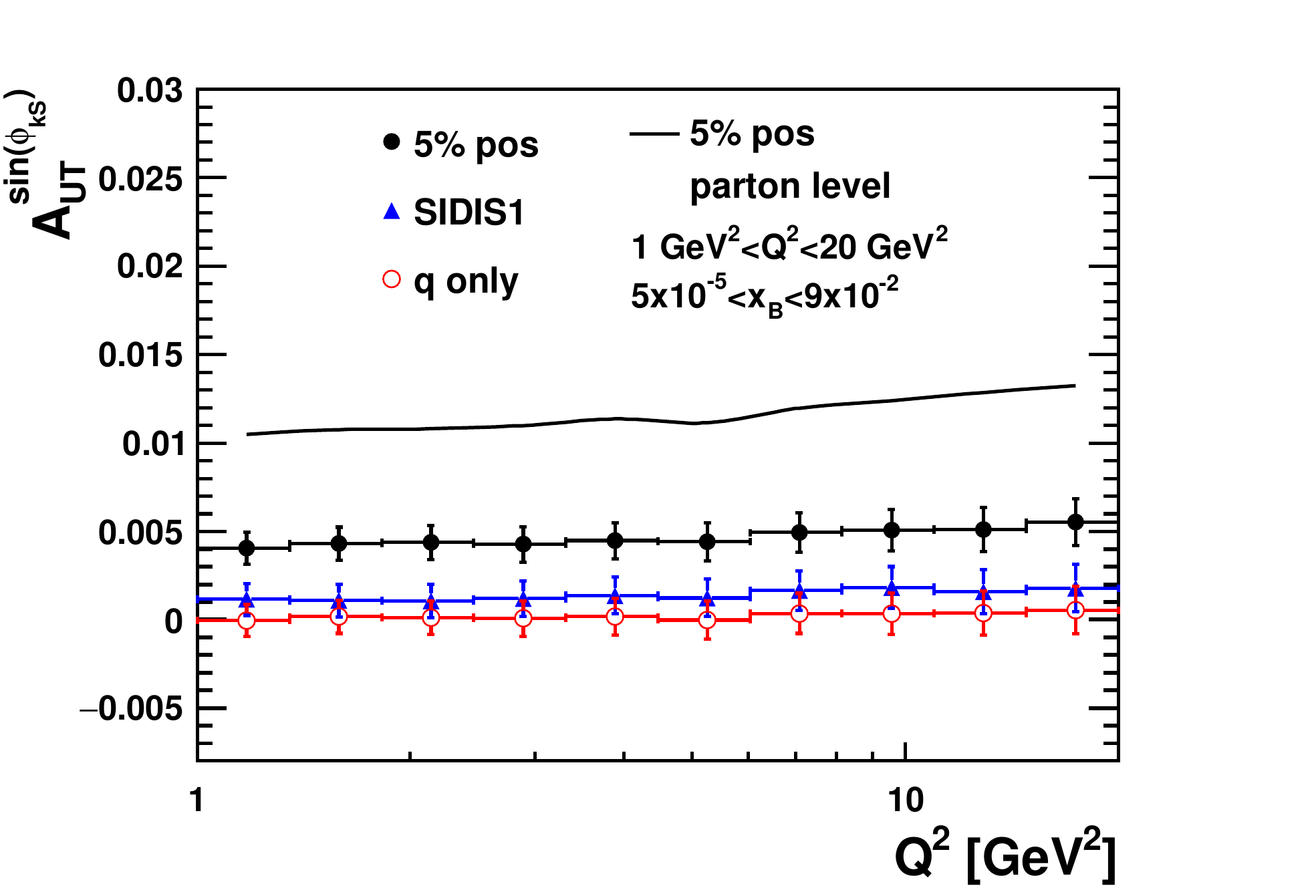}
	}
	\subfigure[]{
		\includegraphics[width=0.45\textwidth]{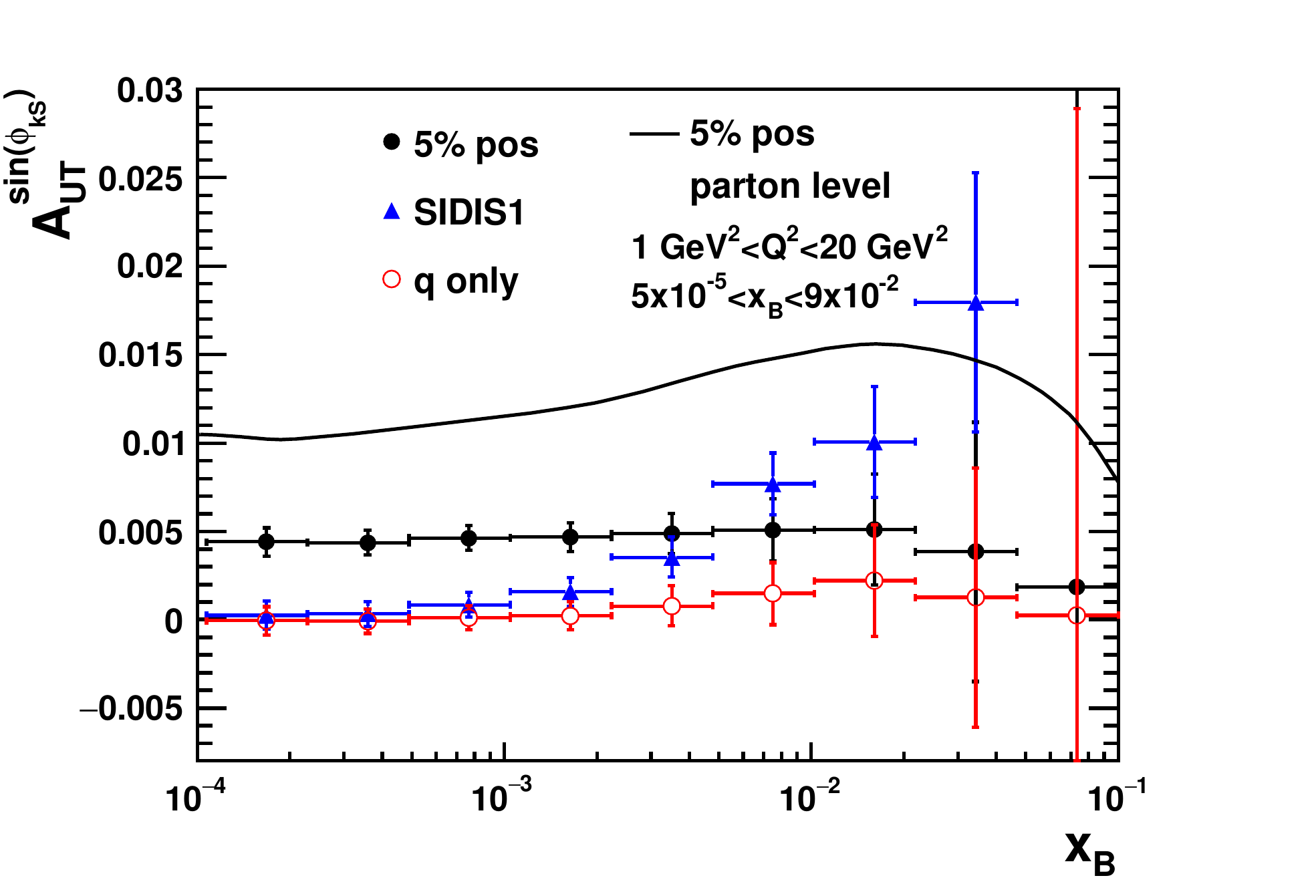}
	}
	\subfigure[]{
		\includegraphics[width=0.45\textwidth]{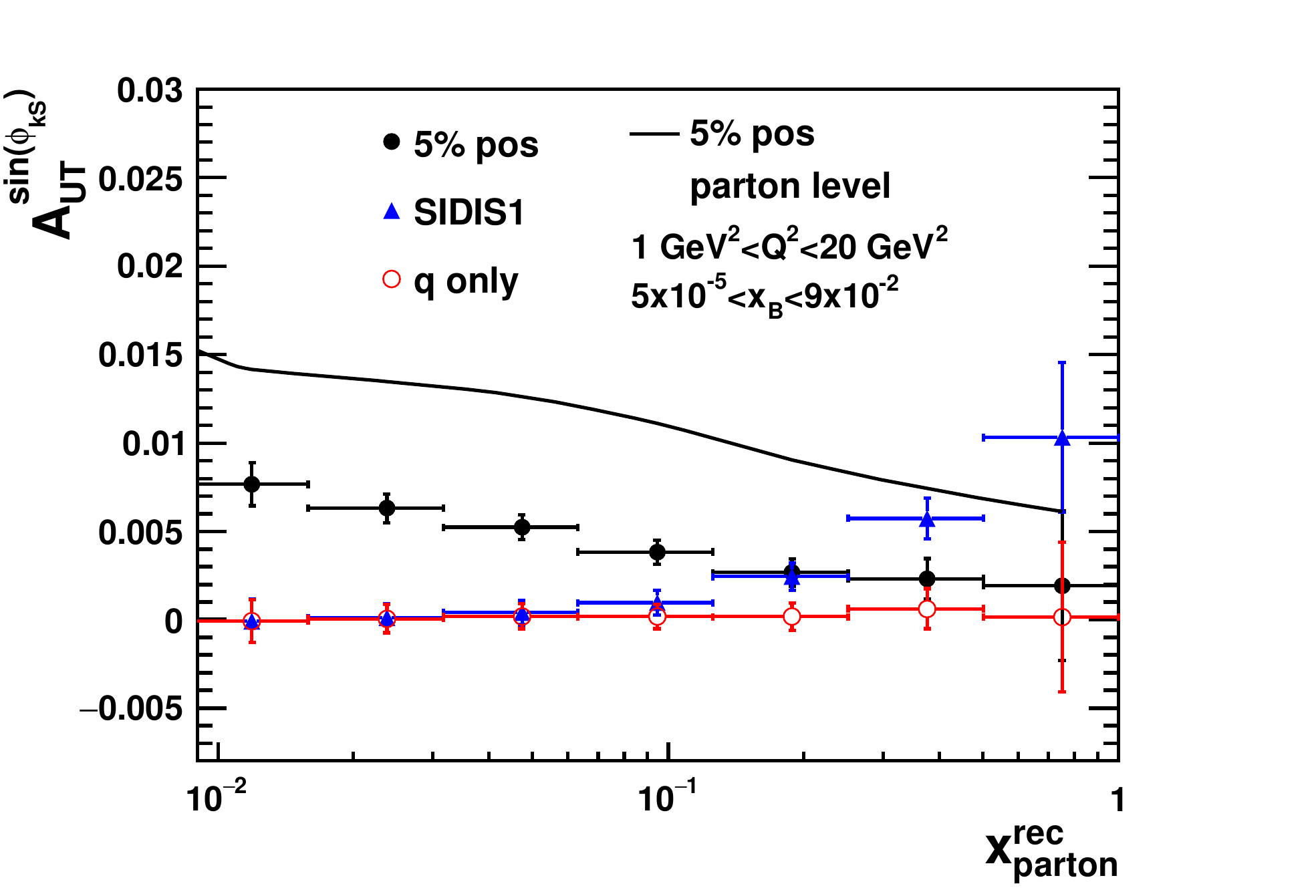}
		\label{fig:dijet_kine_xparton}
	}
	\caption{(Color online) SSA modulation dependent on $Q^{2}$ (a), $x_{B}$ (b) 
		and $\xprec$ (c) for the di-jet channel applying the kinematic cuts: trigger jet
		$p_T^{jet1}>$ 4.5 GeV and associate jet $p_T^{jet2}>$ 4 GeV, $0.01<y<0.95$ and $1\,
		\textrm{GeV}^{2}<Q^{2}<20 \, \textrm{GeV}^{2}$ at the electron-proton beam energy 20
		GeV$\times$ 250 GeV with an integrated luminosity $\mathcal{L}_{int}=10$ fb$^{-1}$.
	}
	\label{fig:dijet_kine_evolve}
\end{figure}
In Fig.~\ref{fig:dijet_kine_evolve}, it is observed that a gluon Sivers function
with the size of 5\% of the positivity bound or the SIDIS1 set can be well
separated from a SSA based on the quark Sivers effect at large $x_{B}$ for an
integrated luminosity of $\mathcal{L}_{int}=10$ fb$^{-1}$. Despite that the
initial parton asymmetry for the dijet process is smaller than that for the di-hadron channel, a larger
fraction of the initial asymmetry survives in the di-jet channel. The shape of
the initial parton level asymmetry is largely preserved in the di-jet asymmetry
in all kinematic variables. This is of great advantage to explore dependence of
the gluon Sivers function on the hard scattering kinematics. Due to the strong
correlation between the momentum of a jet and its mother parton, it is possible
to reconstruct the momentum fraction of the parton participating in the hard
interaction from the di-jet momentum information: \( \xprec
=(p_{T}^{jet1}e^{-\eta^{jet1}}+p_{T}^{jet2}e^{-\eta^{jet2}})/W \). The SSA as a
function of $\xprec$ is shown in Fig.~\ref{fig:dijet_kine_xparton}. The initial
functional form of the  gluon Sivers function on $x$ is well reproduced in the
measured SSA as function of  $\xprec$. The SSA based on a gluon Sivers function
with a magnitude of 5\% of the positivity bound drops while the one based on the
SIDIS1 set increases with $\xprec$. The shape of the initial gluon Sivers
function is largely the same in the measured SSA. This will allow to
distinguish different gluon Sivers models.

\begin{figure}
\centering
\includegraphics[width=0.45\textwidth]{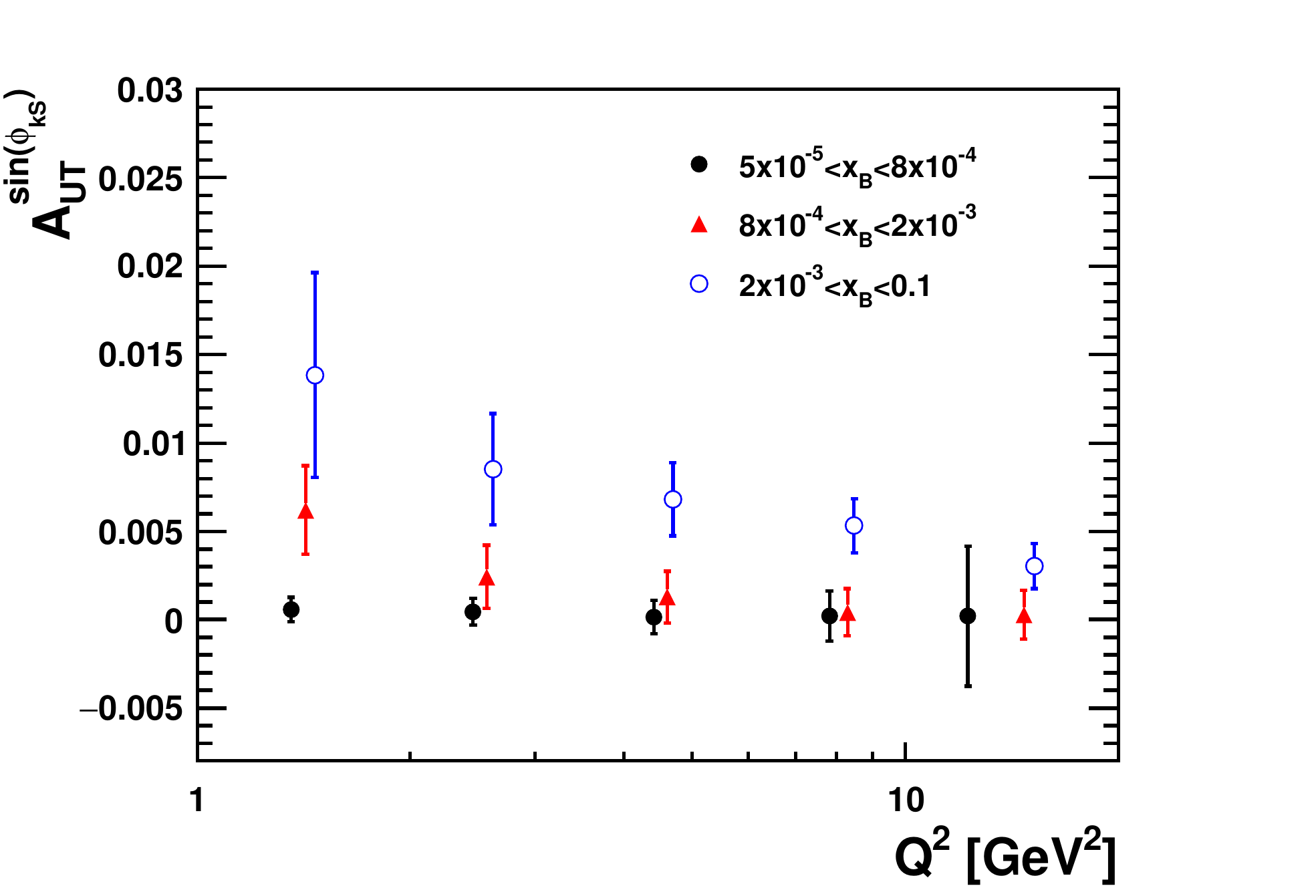}
\caption{(Color online) Sivers modulation for the di-jet channel as function of $Q^2$ 
for three $x_B$ bins using the SIDIS1 gluon Sivers function as input.
}
\label{fig:dijet_evolve_2D}
\end{figure}

With projected high statistics for the di-jet channel, the evolution of the GSF
with $Q^2$ and $x_B$ can be studied utilizing a multi-dimensional binning.
Figure~\ref{fig:dijet_evolve_2D} shows the SSA based on the gluon Sivers
function from the SIDIS1 set as function of $Q^2$ in three $x_B$ bins. The
differential features of the SIDIS1 gluon Sivers function are: the asymmetry
increases in the high $x_B$ bins, which is consistent with the behavior in
$x_B$. The decrease of the SSA as function of $Q^2$ is seen especially in the
high $x_B$ bins. This signature can be utilized to study the evolution of the
gluon Sivers function in the di-jet channel.

\begin{figure}
	\centering
	\subfigure[]{
		\includegraphics[width=0.45\textwidth]{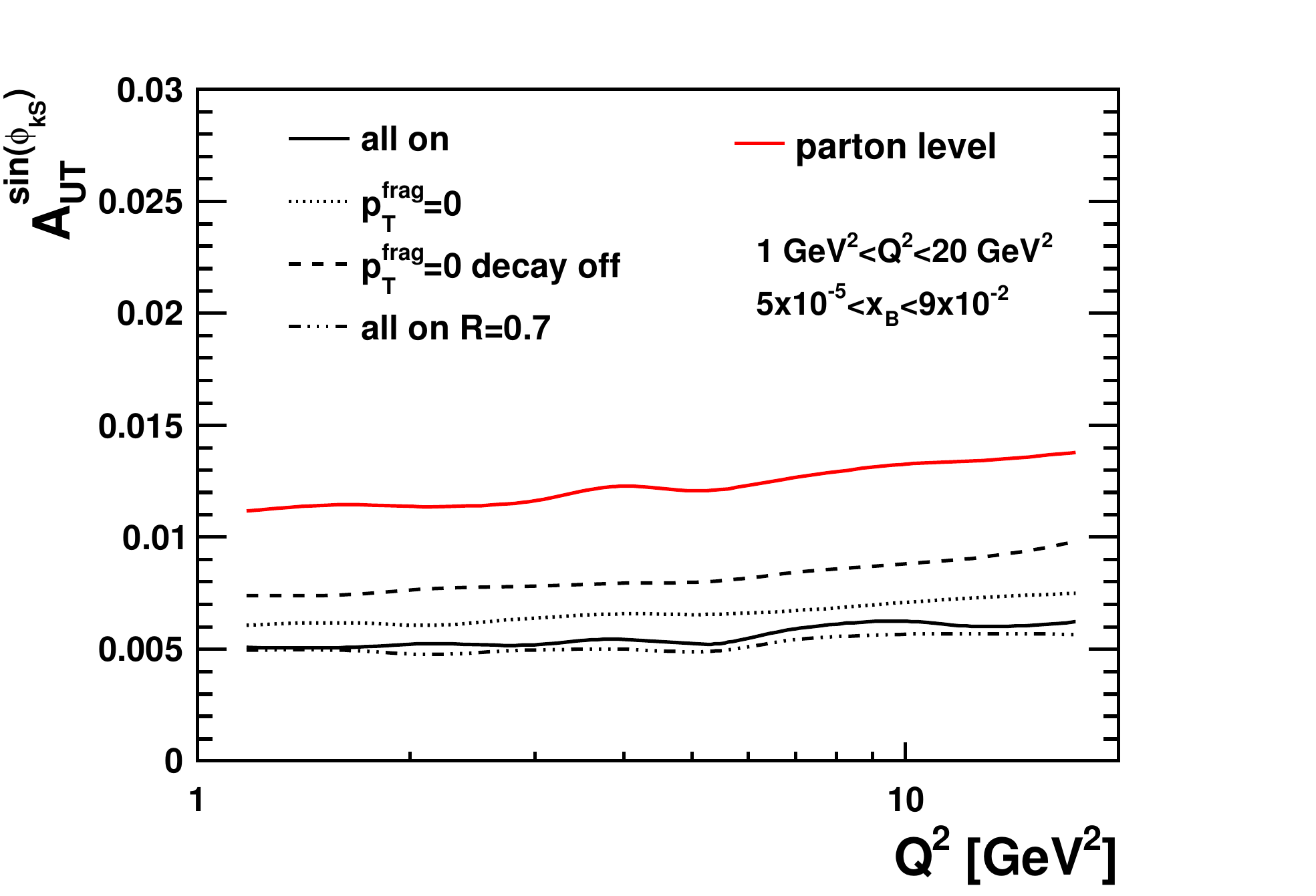}
		\label{fig:dijet_systematic_Q2}
	}

	\subfigure[]{
		\includegraphics[width=0.45\textwidth]{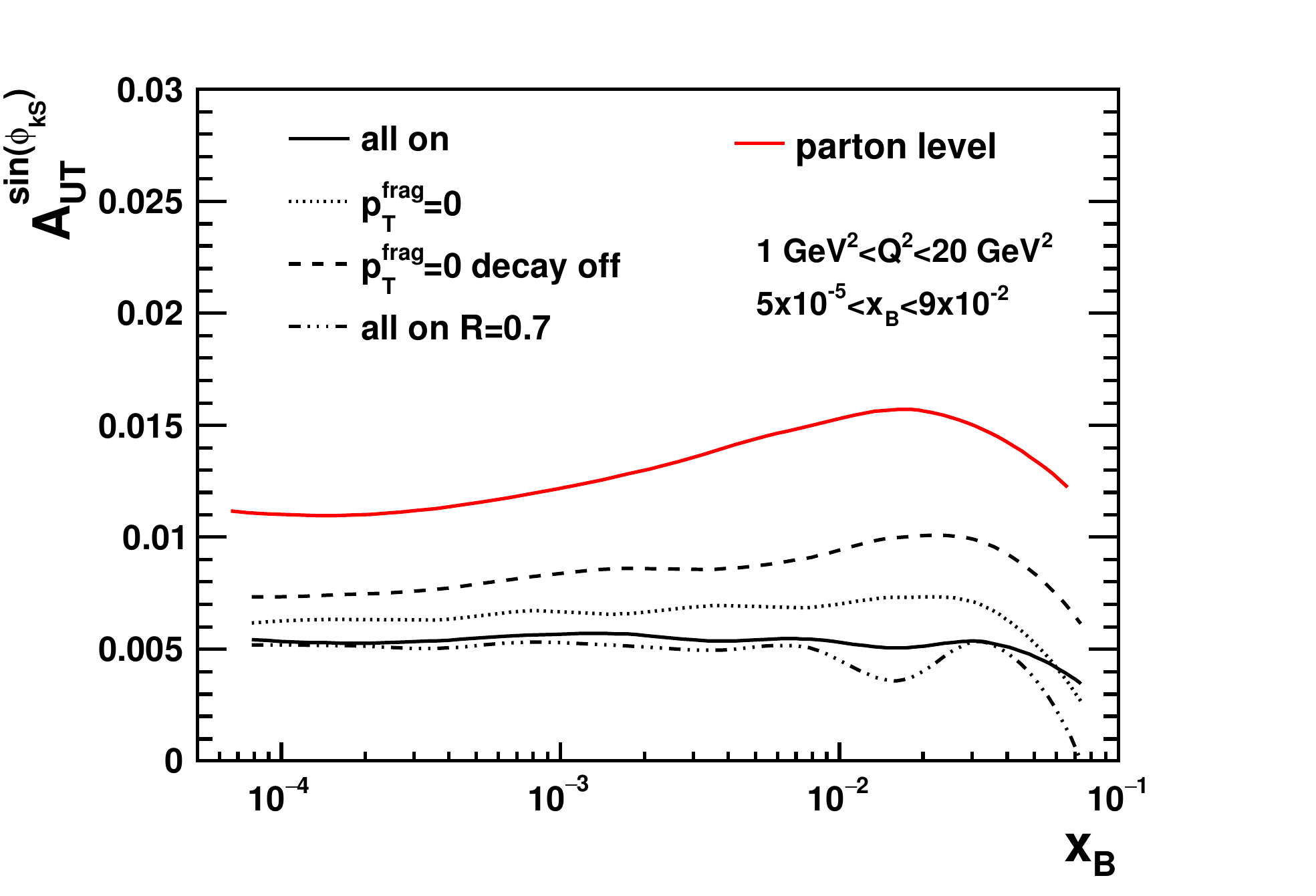}
		\label{fig:dijet_systematic_xB}
	}
\caption{(Color online) Sivers asymmetries for di-jets for different hadronization assumptions. 
Shown are the initial parton level asymmetry (red solid curve), the di-jet asymmetry (black solid curve). The dotted curve represents the di-jet Sivers SSA without fragmentation $p_T$, while the dashed curve represents both no fragmentation $p_T$ and the resonance decays in the fragmentation turned off. The dash-dotted line represents the di-jet asymmetry with a cone size of $R=0.7$. }
	\label{fig:dijet_systematic}
\end{figure}
A more detailed analysis on the main cause of the smearing from the parton level
asymmetry to the measured asymmetry in di-jet production is shown in
Fig.~\ref{fig:dijet_systematic}. We present a comparison of the observed Sivers
asymmetry in di-jet measurement with different hadronization assumptions to the
probed parton level asymmetry. To study the influence of $p_T$ in the
hadronization as well as the effect due to the decay of particle
resonances, we have turned both processes consecutively off in the simulation by
setting the respective PYTHIA parameters (PARJ(21) and MSTJ(21)) to zero. The
solid red and black curves in Fig.~\ref{fig:dijet_systematic} represent the
parton level and measured asymmetry, respectively. Comparing the dotted curve
(fragmentation $p_T$ off) and the dashed curve (particle resonance decay and
fragmentation $p_T$ off) shows clearly the dominant effect is due to resonance
decay in the fragmentation. As is shown in Fig.~\ref{fig:dijet_systematic_xB} of
the asymmetry varying with $x_B$, we observe the resonance decay becomes
slightly more important in the high $x_B$ region. The remaining dilution of the
parton level asymmetry is caused by QCD radiation, the $p_T$ dependence of the
hard scattering process and the ability to measure a small transverse momentum
imbalance with high $p_T$ di-jets. We also perform a study on the impact of
using different algorithms and cone sizes in the jet reconstruction procedure.
It is found that the effect by changing the jet reconstruction algorithm from
anti-$k_T$ to $k_T$ or the cone size from $R=1$ to $R=0.7$ is barely visible. 
We present the result of di-jet asymmetry from cone size $R=0.7$ with the
dash-dotted line in Fig.~\ref{fig:dijet_systematic}. The choice of a smaller
cone size only leads to a slightly smaller asymmetry. It is implied in this
comparison that the di-jet asymmetry size is rather robust in spite of the jet
reconstruction methods.

\begin{figure}
\centering
\includegraphics[width=0.45\textwidth]{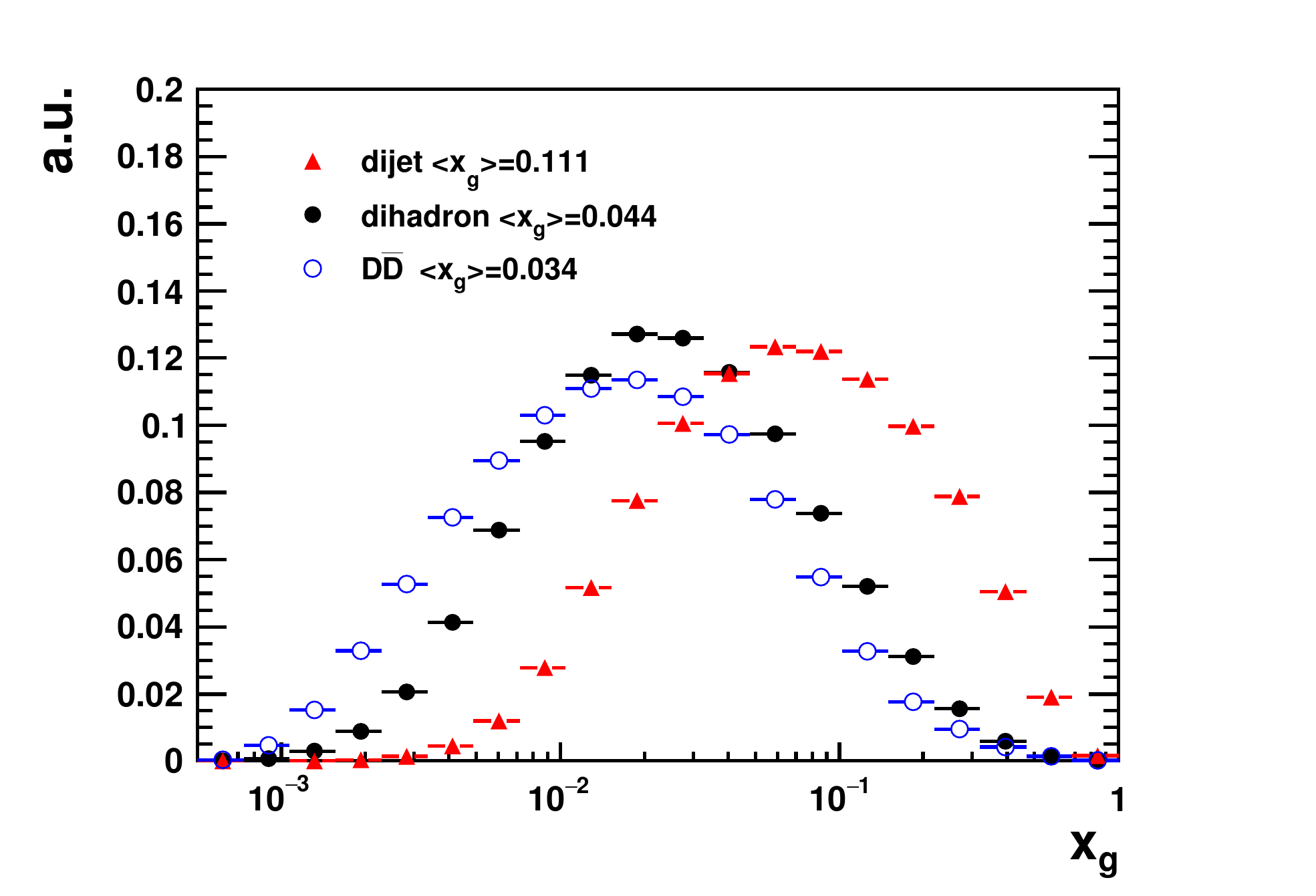}
\caption{(Color online) $x_g$ coverage probed by all the different observables.
	}
	\label{fig:xg_compare}
\end{figure}

An important aspect in comparing the different channels is the coverage in the
gluon momentum fraction $x_g$. Figure~\ref{fig:xg_compare} shows the
$x_g$-distributions for the different channels are complementary. Both the heavy
flavor $D\bar{D}$ mesons and the di-hadrons probe lower values of $x_g$ (
$\langle x_g \rangle \sim 0.03$) and the di-jet channel is probing the larger
$x_g$ range ( $\langle x_g \rangle \sim 0.1$).

\section{Summary}
\label{sec:sum}
We have performed a study on the feasibility of measuring the
gluon Sivers function in high $p_T$ charged di-hadrons, heavy flavor mesons and
di-jet production in polarized $ep^{\uparrow}$ collisions at an EIC. Scanning
different assumptions of the magnitude of the gluon Sivers function provides a
systematic study on the sensitivity of the different channels.

It is found that although the heavy flavor $D\bar{D}$ meson production is the
cleanest channel to tag gluon initiated processes, it is at the same time also
the most statistically challenging process and therefore the sensitivity to
small gluon Sivers effects is limited. An alternative method using inclusive $D$
mesons provides sensitivity to a gluon Sivers function with a magnitude of 10\%
of the positivity bound for a nominal integrated luminosity of
$\mathcal{L}_{int}=10$ fb$^{-1}$. But the smearing between parton level and the
measured asymmetry is significantly increased. The high $p_T$ charged di-hadron
channel is statistically more favorable and can resolve a magnitude of the gluon
Sivers function down to 5\% of the positivity bound. The most precise analyzer
for the gluon Sivers effects at an EIC is the di-jet channel, due to its
statistical advantage it provides the best sensitivity even for the small Sivers
effects and can span the largest $Q^2$-range to study TMD evolution effects. Due
to its tight correlation between parton and jet kinematics, it has the smallest
dilution between the parton level and measured asymmetries. Overall it is thus
the most promising experimental channel to determine and study all features of
the gluon Sivers effect at the future EIC.

Following the classification of the unpolarized gluon TMDs, the gluon Sivers function can also be separated in a WW type and dipole type TMDs depending on the gauge link structure involved in the process. Quark-antiquark production in DIS is probing the WW gluon Sivers function and is related to the one measured in photon pair production in proton-proton collisions through a sign-flip as discussed in Sec.I. It is noted that the dipole type gluon Sivers function can only be accessed through the $p^{\uparrow}p\rightarrow\gamma\ \mathrm{jet}\ X$ process. The future EIC project will play an important role to provide complementary information to further our understanding of the different types of gluon TMDs, in particular for the gluon Sivers function by testing the predicted sign-flip of the WW gluon Sivers function if measured in ep and pp.
\\

\begin{acknowledgments}
We would like to thank M. Diehl, Zhong-Bo Kang for helpful suggestions and
discussions. We are grateful to A. Prokudin for providing us their recent
parameterizations for the quark Sivers function. E.C.A. and J.H.L. acknowledge
the support by the U.S. Department of Energy under contract number DE-SC0012704.
This work was supported by the Fundamental Research Funds for the Central
Universities, China University of Geosciences (Wuhan) No.CUG180615, the National
Key Research and Development Program of China (2016YFE0100900) and the NSFC
(11475068, 11575070).
\end{acknowledgments}

\bibliography{reference}

\end{document}